\documentclass[11pt,letterpaper]{article}

\usepackage{booktabs}

\usepackage{latexsym}
\usepackage{amssymb,amsmath,bm,pgfplots,tikz,bbm}
\pgfplotsset{compat=1.18}
\usepackage{graphicx}
\usepackage{marvosym}
\usepackage{multirow,float}
\usepackage{algpseudocode}
\usepackage{caption}
\usepackage{mathtools}
\usepackage{subcaption}
\usepackage{comment}
\usepackage{enumitem}
\usepackage{xcolor}
\usepackage{microtype}
\allowdisplaybreaks

\usepackage{natbib}
\usepackage{framed}
\usepackage{soul}
\usepackage{longtable}

\usepackage[colorlinks=true,citecolor=blue,linkcolor=black,
urlcolor=blue,bookmarksopen=true,bookmarksnumbered=true,
pdfstartview=FitH,breaklinks=true]{hyperref}
\usepackage{etoolbox}
\makeatletter
\patchcmd{\NAT@citex}
  {\@citea\NAT@hyper@{\NAT@nmfmt{\NAT@nm}\hyper@natlinkbreak{\NAT@aysep\NAT@spacechar}{\@citeb\@extra@b@citeb}\NAT@date}}
  {\@citea\NAT@nmfmt{\NAT@nm}\NAT@aysep\NAT@spacechar\NAT@hyper@{\NAT@date}}{}{}
\patchcmd{\NAT@citex}
  {\@citea\NAT@hyper@{\NAT@nmfmt{\NAT@nm}\hyper@natlinkbreak{\NAT@spacechar\NAT@@open\if*#1*\else#1\NAT@spacechar\fi}{\@citeb\@extra@b@citeb}\NAT@date}}
  {\@citea\NAT@nmfmt{\NAT@nm}\NAT@spacechar\NAT@@open\if*#1*\else#1\NAT@spacechar\fi\NAT@hyper@{\NAT@date}}{}{}
\makeatother

\usepackage{dcolumn}
\newcolumntype{.}{D{.}{.}{-1}}
\newcolumntype{d}[1]{D{.}{.}{#1}}

\usepackage{theorem}
\theoremstyle{plain}
\theoremheaderfont{\normalfont\scshape}

\newtheorem{proposition}{Proposition}
\newtheorem{assumption}{Assumption}

\newtheorem{lemma}{Lemma}

\newcommand{\qed}{\hfill \ensuremath{\Box}}
\newcommand{\indep}{\mbox{$\perp\!\!\!\perp$}}

\renewcommand{\d}{\mathrm{d}}

\DeclareMathOperator\Cov{Cov}
\DeclareMathOperator*{\argmin}{argmin}

\newcommand{\norm}[1]{\left\lVert#1\right\rVert}
\newenvironment{proof}{\vspace{1ex}\noindent{\bf Proof}\hspace{0.5em}}
{\hfill\qed\vspace{1ex}}
\usepackage{kantlipsum}
\allowdisplaybreaks
\usepackage{rotating}
\usepackage{arydshln}
\usepackage{threeparttable}
\usepackage[compact]{titlesec}

\usepackage[ruled,linesnumbered,vlined]{algorithm2e}
\usepackage{pifont}

\newcommand\E{\mathbb{E}}

\renewcommand\P{\mathbb{P}}

\newcommand\V{\mathbb{V}}

\newcommand\bx{\bm{x}}
\newcommand\bX{\bm{X}}

\usetikzlibrary{decorations.markings,decorations.pathmorphing,
shapes.geometric,arrows,backgrounds,positioning,fit,matrix,shapes,
decorations,calc,arrows.meta}
\tikzset{auto,node distance=1 cm and 1 cm,semithick,
state/.style={circle,draw,minimum width=0.7 cm},
point/.style={circle,draw,inner sep=0.04cm,fill,node contents={}},
bidirected/.style={Latex-Latex,dashed},
el/.style={inner sep=2pt,align=left,sloped}}

\graphicspath{{figs/}}
\hypersetup{pdftitle={A General Statistical Framework for Attention and Manipulation Checks},
pdfauthor={Kentaro Nakamura and Kosuke Imai}}

\begin{document}
\title{\bf A General Statistical Framework for Attention \\ and Manipulation Checks\thanks{We thank Lea Dinsing, Max Goplerud, and Libby Jenke for helpful discussions. }}
\author{Kentaro Nakamura\thanks{Ph.D. Student, John F. Kennedy School
    of Government, Harvard University. Email:
    \href{mailto:knakamura@g.harvard.edu}{knakamura@g.harvard.edu}.} \and
    Kosuke Imai\thanks{Edith and Benjamin Geisinger
    Professor of Government and of Statistics,
    Harvard University, Cambridge, MA 02138. Phone: 617--384--6778, Email:
    \href{mailto:Imai@Harvard.Edu}{Imai@Harvard.Edu}, URL:
    \href{https://imai.fas.harvard.edu}{https://imai.fas.harvard.edu}.}}
\date{\today}
\maketitle\thispagestyle{empty}

\begin{abstract}
Attention and manipulation checks are widely used in survey experiments to assess respondents' attentiveness and engagement with treatment, respectively. Yet these checks are subject to measurement error, which may itself depend on treatment when checks are administered after treatment exposure. Restricting analyses to respondents who pass these checks can therefore introduce substantial bias. We propose a general statistical framework for analyzing survey experiments with pre- or post-treatment attention and manipulation checks while accounting for measurement error. We show that three separate checks suffice to identify the proportion of respondents who would be attentive or successfully manipulated under treatment, provided that the checks are conditionally independent among treated respondents given their true, unobserved attentiveness or manipulation status and pre-treatment covariates. Under additional mild assumptions, this result also identifies the average treatment effect among respondents who would be attentive or successfully manipulated under treatment. Building on these identification results, we propose an estimator that allows flexible covariate modeling and establish its asymptotic normality. Applications to framing and digital media literacy experiments demonstrate how accounting for measurement error yields treatment effect estimates that are substantively different from those obtained using conventional approaches.
\end{abstract}


\newpage
\section{Introduction}
\label{sec:introduction}

In survey experiments, random assignment of treatment identifies the intention-to-treat (ITT) effect, but this estimand does not distinguish respondents who pay attention to or engage with the treatment from those who do not.
To assess attentiveness and treatment engagement, researchers commonly administer attention or manipulation checks to determine, for example, whether respondents carefully read instructions or understood the intended manipulation \citep[e.g.,][]{berinsky2014separating,kane2019checking}.
Although informative, these checks are imperfect measures of respondents' underlying attentiveness or engagement. 
Some inattentive respondents may pass a check by chance, whereas successfully manipulated respondents may fail because they misunderstand the manipulation check questions.
In addition, when checks are administered after treatment assignment, treatment itself may also affect measurement error \citep[e.g.,][]{montgomery2018conditioning,aronow2019note,blackwell2025priming}. 
Therefore, restricting the analysis to respondents who pass these checks can introduce substantial bias.

These challenges raise a fundamental question: how can researchers use imperfect attention and manipulation checks to estimate treatment effects among respondents who would be attentive or successfully manipulated under treatment?
We propose a general statistical framework for estimating this quantity, which we call the engaged average treatment effect (eATE), where engagement refers to attentiveness or successful manipulation, depending on the application.
Our framework accommodates both pre-treatment engagement, which is unaffected by treatment assignment, and post-treatment engagement, which may depend on assignment.
We show that the eATE can be identified using three potentially imperfect checks that are conditionally independent given true, unobserved engagement status, treatment assignment, and pre-treatment covariates.

Building on this identification result, we develop an estimation procedure that accommodates flexible covariate modeling and provides valid uncertainty quantification.
Specifically, we use spline approximations to allow both engagement probabilities and the measurement qualities of each check to smoothly vary with respondent characteristics.
We then propose a cross-fitted estimator with a valid asymptotic confidence interval that accounts for uncertainty in fitting the measurement model, leveraging the results from the semiparametric statistics literature \cite[e.g.,][]{tsiatis2006semiparametric,chernozhukov2018double}.
Lastly, we also develop a statistical test that can be used to diagnose the potential violations of the relevance conditions under the conditional independence assumptions. 

We apply the proposed methodology to two published experiments, one with attention checks and the other with post-treatment manipulation checks.
In the civil liberties framing experiment of \citet{berinsky2014separating}, our approach yields a larger estimated effect on rally approval among attentive respondents than either the full-sample ITT estimate or the estimate among respondents who pass the observed attention checks.
In the digital media literacy experiment of \citet{guess2020digital}, we estimate that the proportion of respondents who would be successfully manipulated exceeds the proportion who pass all three manipulation checks.
The resulting eATE estimates are smaller in absolute magnitude than the instrumental variable estimates but larger than the ITT estimates.
Together, these applications demonstrate how accounting for measurement error in attention and manipulation checks can materially change the estimated magnitude of treatment effects.

\paragraph{Related literature.}
This paper contributes to the literature on attention and manipulation checks by developing a new methodological framework that accommodates both pre- and post-treatment checks with measurement error. 
Although these checks are widely used in survey experiments across the social science, there exist few statistical methods that properly account for measurement error in these checks.

For attention checks, administering them before treatment avoids post-treatment selection bias, but the checks themselves may prime respondents and alter their subsequent responses.
In contrast, restricting the analysis to respondents who pass a post-treatment check may introduce selection bias
\citep{montgomery2018conditioning,aronow2019note}.
These concerns have motivated partial identification approaches to post-treatment selection \citep{aronow2019note}, imperfect screening \citep{tyler2024statistical,tyler2025bounding}, and the trade-off between priming and post-treatment bias when measuring moderators \citep{blackwell2025priming}.
Unlike these partial identification approaches, we propose a unified methodological framework that provides point identification of the eATE under the key conditional independence assumption regardless of whether checks are done before or after treatment.
Our methodology also quantifies estimation uncertainty, providing asymptotically valid confidence intervals for the eATE estimates.

\citet{mathur2025statistical_a} takes a different approach, treating attention checks as tools for identifying respondents whose responses are free of measurement error.
The author proposes adjusting for covariates that affect both attentiveness and the outcome when analyzing respondents who pass the checks. 
This approach, however, assumes that passing the attention checks accurately identifies attentive respondents.
In contrast, our methodology accounts for measurement error that is inherent in attention checks.

Accounting for manipulation checks is more challenging because successful manipulation can only be assessed after treatment assignment.
\cite{aronow2019note} and \cite{blackwell2025priming} apply partial identification approaches to this post-treatment selection problem.
Others, including \cite{mathur2025statistical_m}, propose the use of instrumental variables to estimate the eATE where randomized treatment assignment serves as an instrument for the observed manipulation status.
However, this approach requires the exclusion restriction that treatment assignment affects the outcome only through the observed manipulation status, essentially precluding the possibility of any measurement error in manipulation checks and the existence of alternative causal pathways.
In contrast, our approach does not require any individual manipulation check to perfectly measure the underlying manipulation status and allows treatment assignment to affect the outcome directly among respondents who would be manipulated under both treatment conditions.

More broadly, \citet{hazlett2026post} study the eATE in the context of post-treatment selection, which they call the treatment reactive average causal effect (TRACE).
The authors propose a partial identification approach that allows for a nonzero average treatment effect among respondents who would not react to the treatment.
However, their approach does not explicitly account for measurement error in the observed post-treatment variable.
In contrast, our methodology accounts for measurement error in manipulation checks and provides point identification of the eATE, by assuming a zero average treatment effect among respondents who would not be manipulated under either treatment condition.

Finally, the proposed methodology builds on the literature on measurement error and proxy-based identification, which exploits multiple imperfect measurements of a latent variable to identify causal effects under the assumptions of non-differential measurement error. 
Such approaches have been developed for unobserved confounders \citep{kuroki2014measurement, miao2018identifying, tchetgen2024introduction,cui2024semiparametric}, unobserved treatments \citep{zhou2024causal}, and latent outcomes \citep{guo2026proximal}, with a wide range of applications, including unobserved demographic characteristics \citep{mccartan2025estimating,chasalow2026probabilistic} and text annotations \citep{nakamura2025surrogate,egami2026debiased}.
We extend this approach to the setting of attention and manipulation checks in randomized experiments, where respondents' underlying attentiveness or engagement is latent and each observed check may contain  measurement error.

\paragraph{Organization of the paper.}
The rest of this paper proceeds as follows.
Section~\ref{sec:motivation} introduces two empirical applications. 
Section~\ref{sec:methodology} presents the identification results. Section~\ref{sec:estimation} then develops estimation and inference procedures. Section~\ref{sec:testing} discusses a diagnostics test regarding the key identification assumption.
Section~\ref{sec:simulation} then confirms the performance on the simulated data, and Section~\ref{sec:applications} applies the proposed methodology to the motivating examples introduced earlier.
Finally, Section~\ref{sec:conclusion} concludes by discussing directions for future research.

\section{Motivating Empirical Applications}
\label{sec:motivation}

This section introduces two motivating applications: one uses attention checks to assess respondents' attentiveness, and the other uses manipulation checks to evaluate their engagement with the treatment stimulus.
Although the checks are administered both before and after treatment in the first application and only after treatment in the second, our proposed methodology accommodates both settings, provided that at least three well-designed checks are available (see Section~\ref{sec:methodology}).
We revisit both applications in Section~\ref{sec:applications}, where we analyze their data using the proposed methodology.

\subsection{Civil Liberties Framing Experiment with Attention Checks}
\label{example:attention}

A central question in the study of public opinion is whether people's preferences depend on how political issues are presented.
In a classic study, \citet{nelson1997media} examined this question through a survey experiment on civil liberties.
Respondents read an article about a proposed Ku Klux Klan (KKK) rally that presented one of two frames.
The free speech frame emphasized the KKK's right to express its views and the public's right to hear them, whereas the public safety frame emphasized the potential for violence and threats to public order.
The authors found that respondents exposed to the free speech frame were more willing to allow the rally than those exposed to the public safety frame. 
This finding suggests that news framing can shape political tolerance by emphasizing different considerations.

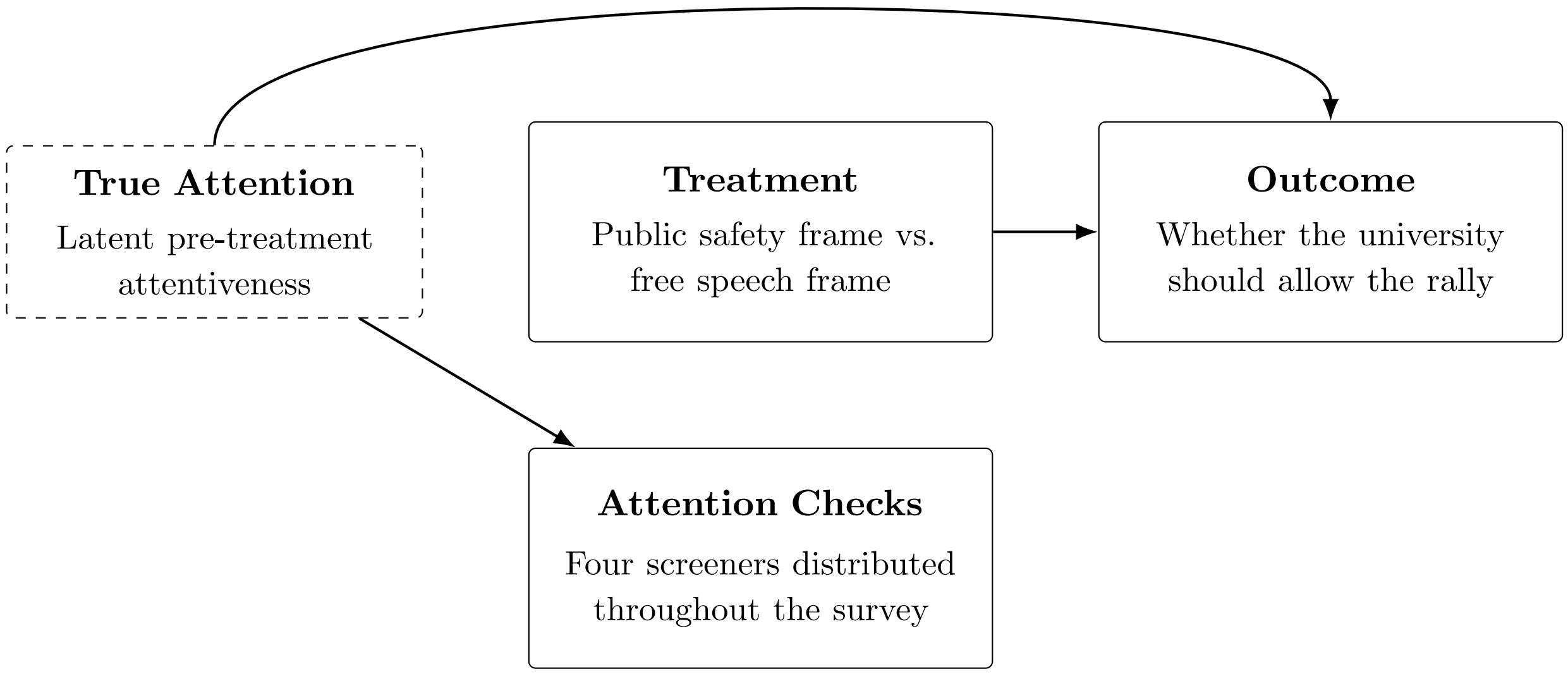
\begin{figure}[t]
\centering
\begin{tikzpicture}[
    >=Latex,
    proc/.style={
        draw, rounded corners=2pt,
        align=center,
        text width=4.5cm,
        minimum height=2.3cm,
        inner sep=5pt
    },
    latent/.style={
        draw, dashed, rounded corners=2pt,
        align=center,
        text width=4.0cm,
        minimum height=1.8cm,
        inner sep=5pt
    },
    node distance=1.1cm
]
    \node[latent] (attention) {
        \textbf{True Attention}\\[4pt]
        \small Latent pre-treatment\\
        attentiveness
    };
    
    \node[proc, right=of attention] (treat) {
        \textbf{Treatment}\\[4pt]
        \small Public safety frame vs.\\
        free speech frame
    };

    \node[proc, right=of treat] (outcome) {
        \textbf{Outcome}\\[4pt]
        \small Whether the university\\
        should allow the rally
    };

    \node[proc, below=of treat] (check) {
        \textbf{Attention Checks}\\[4pt]
        \small Four screeners distributed throughout the survey
    };

    \draw[->, thick] (treat) -- (outcome);
    \draw[->, thick]
    (attention.north)
    .. controls +(0,1.9cm) and +(0,1.5cm) ..
    (outcome.north);
    \draw[->, thick] (attention) -- (check);

\end{tikzpicture}
\caption{Experimental design and attention checks in the
civil liberties framing experiment of
\cite{berinsky2014separating}. Respondents read a news
article emphasizing either public safety or free speech.
They then report whether the university should allow the
KKK rally. The authors measured the attentiveness using attention checks.}
\label{fig:procedure1}
\end{figure}

Although the original study found a framing effect, respondent inattention may attenuate the observed effect.
To explore this possibility, \citet{berinsky2014separating} replicated the civil liberties framing experiment using multiple attention checks.
Figure~\ref{fig:procedure1} illustrates the experimental design.
The authors administered four attention checks throughout the survey, randomizing their order for each respondent.
The checks asked about respondents' favorite color, current feelings, website preferences, and political interest.
Rather than answering these questions directly, respondents were instructed to select specific options to demonstrate that they had read the instructions carefully.
For example, the favorite-color question instructed respondents to select both red and green, regardless of their actual preferences.
The first screener was administered before the framing
experiment, whereas the remaining three were administered
afterwards.

\begin{table}[t]
\centering
\begin{minipage}[t]{0.65\textwidth}
\centering
\textbf{(a) Pearson correlations}\par\medskip
\begin{tabular}{lcccc}
\toprule
 & Web & Interest & Color & Feeling \\
\midrule
Web      & 1 & 0.47 & 0.41 & 0.41 \\
Interest &   & 1    & 0.46 & 0.42 \\
Color    &   &      & 1    & 0.44 \\
Feeling  &   &      &      & 1    \\
\bottomrule
\end{tabular}
\end{minipage}%
\hfill
\begin{minipage}[t]{0.30\textwidth}
\centering
\textbf{(b) Pass rates}\par\medskip
\begin{tabular}{lc}
\toprule
Screener & Pass rate \\
\midrule
Web      & 0.69 \\
Interest & 0.73 \\
Color    & 0.81 \\
Feeling  & 0.81 \\
\bottomrule
\end{tabular}
\end{minipage}

\caption{Pairwise Pearson correlations and marginal pass rates
for the four attention checks used in the experiment of
\citet{berinsky2014separating} (based on the civil liberties
framing experiment, Wave 2 of Study 1).
Web, Interest, Color, and Feeling refer to attention checks
that ask about respondents' website preferences, political
interest, favorite color, and current feelings, respectively.
Each attention check is coded as one if the respondent
correctly follows the instructions and zero otherwise.}
\label{tab:attention_correlation}
\end{table}

This example highlights an important limitation of attention checks.
As \citet{berinsky2014separating} acknowledge, these checks are imperfect measures of respondents' underlying attentiveness.
Inattentive respondents may pass by chance, whereas attentive respondents may fail because they misread the instructions or make a response error.
Indeed, Table~\ref{tab:attention_correlation} shows only modest correlations and varying pass rates across respondents' performance on individual checks.
Therefore, respondents who pass the checks need not be representative of truly attentive respondents.
Without additional restrictions on the measurement error process, the resulting bias can be large and operate in either direction.

Although \citet{berinsky2014separating} recommend using multiple checks, aggregating multiple imperfect measures does not, by itself, resolve the measurement error problem.  
In their analysis, the authors count the number of checks each respondent passes and report treatment effects separately for each count.
They find that the estimated effect is attenuated for those who passed 0 or 1 check when compared to those who passed all checks.
However, it is unclear whether this difference reflects a true difference in attentiveness or other sources of heterogeneity, such as differences in respondents' unobserved characteristics that are correlated with both their attentiveness and their treatment effect.

This study calls for a statistical framework that accounts for measurement error problems of attention checks while allowing for heterogeneity in treatment effects across respondents with different levels of attentiveness.

\subsection{Digital Media Literacy Experiment with Manipulation Checks}\label{example:manipulation}

The spread of misinformation on social media has raised growing concerns about citizens' ability to evaluate political information \citep{allcott2017social,lazer2018science}.
One potential solution is to improve individuals' digital media literacy.
To investigate the effectiveness of this approach, \citet{guess2020digital} examine whether a brief digital media literacy intervention improves people's ability to distinguish mainstream news from false news.
In this experiment, respondents are randomly assigned to the treatment condition that asks them to read ten tips from Facebook's campaign for identifying false news or to the control condition that provides no tips.
The tips encourage readers to scrutinize web addresses and news sources, examine supporting evidence, and compare reports across outlets. 
Respondents subsequently rate the accuracy of news headlines on a four-point scale. 
The outcome of interest is the difference between respondents' perceived accuracy ratings for mainstream and false news headlines.

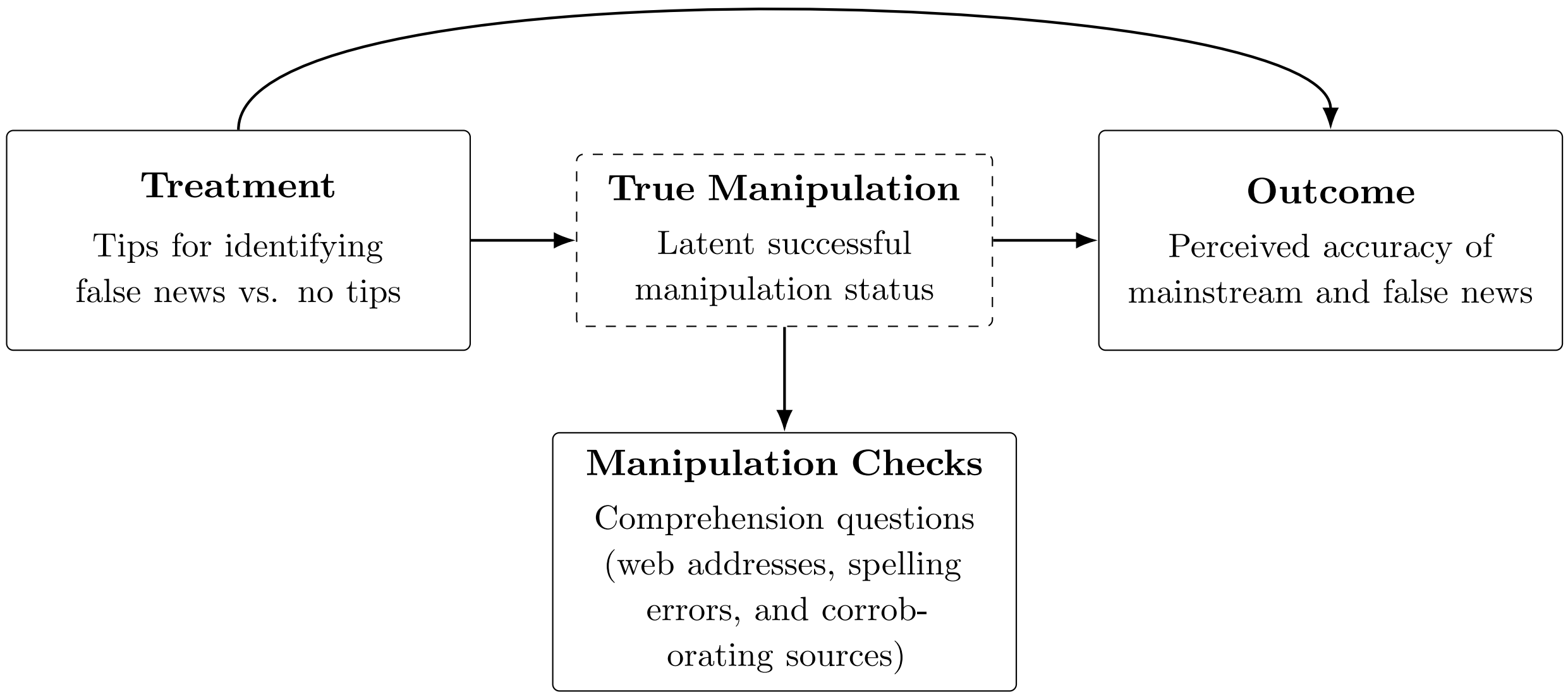
\begin{figure}[t]
\centering
\begin{tikzpicture}[
    >=Latex,
    proc/.style={
        draw, rounded corners=2pt,
        align=center,
        text width=4.5cm,
        minimum height=2.3cm,
        inner sep=5pt
    },
    latent/.style={
        draw, dashed, rounded corners=2pt,
        align=center,
        text width=4.0cm,
        minimum height=1.8cm,
        inner sep=5pt
    },
    node distance=1.1cm
]
    \node[proc] (treat) {
        \textbf{Treatment}\\[4pt]
        \small Tips for
        identifying false news
        vs. no tips
    };

    \node[latent, right=of treat] (manip) {
        \textbf{True Manipulation}\\[4pt]
        \small Latent successful\\
        manipulation status
    };

    \node[proc, right=of manip] (outcome) {
        \textbf{Outcome}\\[4pt]
        \small Perceived accuracy of\\
        mainstream and false news
    };

    \node[proc, below=of manip] (check) {
        \textbf{Manipulation Checks}\\[4pt]
        \small Comprehension questions\\
        (web addresses, spelling errors,
        and corroborating sources)
    };

    \draw[->, thick] (treat) -- (manip);
    \draw[->, thick] (manip) -- (outcome);
    \draw[->, thick] (manip) -- (check);

    \draw[->, thick]
        (treat.north)
        .. controls +(0,1.8cm) and +(0,1.5cm) ..
        (outcome.north);

\end{tikzpicture}
\caption{Experimental design and manipulation checks in the
digital media literacy experiment of \citet{guess2020digital}.
Respondents receive the digital media literacy intervention
and evaluate the accuracy of mainstream and false news
headlines. True manipulation is modeled as a latent
post-treatment variable that affects respondents' outcomes
and their performance on the observed manipulation checks.
The direct arrow from treatment to outcome allows treatment
effects beyond those operating through manipulation status.}
\label{fig:procedure2}
\end{figure}

\begin{table}[t]
\centering
\begin{minipage}[t]{0.60\textwidth}
\centering
\textbf{(a) Pearson correlations}\par\medskip
\begin{tabular}{lccc}
\toprule
 & URLs & Misspellings & Sources \\
\midrule
URLs         & 1 & 0.37 & 0.31 \\
Misspellings &   & 1    & 0.29 \\
Sources      &   &      & 1    \\
\bottomrule
\end{tabular}
\end{minipage}%
\hfill
\begin{minipage}[t]{0.35\textwidth}
\centering
\textbf{(b) Pass rates}\par\medskip
\begin{tabular}{lc}
\toprule
Check & Pass rate \\
\midrule
URLs         & 0.824 \\
Misspellings & 0.824 \\
Sources      & 0.821 \\
All three    & 0.656 \\
\bottomrule
\end{tabular}
\end{minipage}

\caption{Pairwise Pearson correlations and pass rates
for the three manipulation checks used in the digital
media literacy experiment of \citet{guess2020digital}.
URLs, Misspellings, and Sources refer to checks assessing
respondents' understanding of misleading web addresses,
spelling errors, and corroboration by multiple trusted
sources, respectively.
Each check is coded as one if the respondent correctly
answers the question on the first attempt and zero otherwise.
The last row of panel (b) reports the proportion of
respondents who passed all three checks.}
\label{tab:manipulation_pass_rates}
\end{table}

A fundamental challenge in this study is that treatment assignment does not necessarily guarantee successful intervention.
The effectiveness of the intervention depends on whether respondents pay attention to the tips and understand the advice. 
To assess whether respondents successfully receive the intervention, \citet{guess2020digital} administer three manipulation checks in the form of comprehension questions that cover three aspects of the ten tips after the intervention (see Figure~\ref{fig:procedure2} for the illustration of the experimental design).
The authors define successful manipulation as correctly answering all three questions on the first attempt.
However, as shown in Table~\ref{tab:manipulation_pass_rates}, these checks are only moderately correlated.
Moreover, even though the pass rate for each check is relatively high, only 66\% of treated respondents passed all three checks.


In addition to potential measurement error in the observed manipulation checks, this example highlights additional challenges of using post-treatment checks. 
Since the checks are conducted after treatment assignment, subgroup analyses restricted to respondents who pass the checks may suffer from post-treatment selection bias \citep{montgomery2018conditioning, aronow2019note}. 
As a result, a common approach in this context is to use instrumental variables \citep[e.g.,][]{mathur2025statistical_m}. However, the validity of the instrumental variable approach relies on the exclusion restriction, which may be difficult to justify in many applications, including this study. 
For example, respondents who fail to correctly answer certain comprehension questions may nevertheless understand and apply some of the tips, improving their ability to distinguish mainstream news from false news. 
Therefore, treatment assignment may affect the outcome even among respondents classified as failing the manipulation checks, violating the exclusion restriction. 

This study motivates the development of a statistical framework that accounts for imperfect measurement of manipulation status while avoiding the full exclusion restriction required by the standard instrumental variable approach.

\section{The Proposed Framework}
\label{sec:methodology}

In this section, we introduce the proposed methodology for estimating the engaged average treatment effect (eATE) in survey experiments with imperfect attention or manipulation checks.
We begin by defining the eATE as the causal quantity of interest and then establish its identification under the key conditional independence assumption. 
We then briefly discuss how the proposed methodology relaxes the exclusion restriction assumption required by the standard instrumental variable approach. 

\subsection{The engaged Average Treatment Effect (eATE)}
\label{sec:setup}

Consider a simple random sample of $n$ respondents from a population of interest. 
For each respondent $i=1,\ldots,n$, let $T_i\in\mathcal{T}=\{0,1\}$ denote the treatment assignment, where $T_i=1$ indicates treatment and $T_i=0$ indicates the control condition. 
We may also observe the pre-treatment covariates, which we denote $\bX_i$ when available. 
Let $Y_i(t)$ denote the potential outcome under treatment assignment $t\in\mathcal{T}$. 
We use $Y_i=Y_i(T_i)$ to denote the observed outcome.

Our causal quantity of interest is the average treatment effect among respondents who are attentive or successfully manipulated, which we call the engaged average treatment effect (eATE).
Because our framework encompasses both attention and manipulation checks, we use the term \emph{engagement} to refer to attention or successful manipulation, depending on the application. 

Formally, let $W_i(t)\in\{0,1\}$ denote the potential engagement status under treatment assignment $t$, where $W_i(t)=1$ indicates that respondent $i$ would be engaged under this assignment. 
We denote the realized engagement status by $W_i=W_i(T_i)$.
This notation accommodates both pre-treatment attention and post-treatment manipulation. 
When engagement represents pre-treatment attention, treatment assignment does not affect engagement status so that $W_i(1)=W_i(0)=W_i$ for every respondent. 
In contrast, when engagement represents post-treatment manipulation, engagement status may depend on treatment assignment, allowing $W_i(1) \neq W_i(0)$ for some respondents.

We begin by making two standard assumptions about survey experiments.  
First, we assume consistency, requiring that each respondent's potential outcomes and engagement status depend only on their own treatment assignment. 
This assumption rules out spillover effects between respondents.
\begin{assumption}[Consistency]\label{ass:consistency}
The observed outcome and realized engagement status satisfy
\begin{align*}
    Y_i=Y_i(T_i)\quad \text{and} \quad W_i=W_i(T_i), \quad \text{for all } i=1,\ldots,n.
\end{align*}
\end{assumption}

Second, we assume that treatment is randomly assigned, independently of all potential quantities conditional on pre-treatment covariates.
This accommodates a variety of settings, including complete and stratified randomization.
\begin{assumption}[Randomization of Treatment Assignment]
\label{ass:randomization}
Treatment assignment satisfies
\begin{align*}
    T_i \ \indep \
    \{Y_i(1),Y_i(0), W_i(1),W_i(0)\} \mid \bX_i,
\end{align*}
with the known assignment probability $\pi_t(\bx) := \Pr(T_i=t \mid \bX_i = \bx )>0$ for all $\bx$ and $t\in\mathcal{T}$.
\end{assumption}

Under Assumptions~\ref{ass:consistency}~and~\ref{ass:randomization}, the average treatment effect (ATE) is identified by the mean difference in outcome between the treatment and control groups with necessary covariate adjustment.
Our primary quantity of interest, however, is the average treatment effect among respondents who would be engaged under treatment, or the engaged average treatment effect (eATE), which we define as:
\begin{align}
    \Psi
    &:=\E[Y_i(1)-Y_i(0)\mid W_i(1)=1]. \label{target}
\end{align}
When engagement represents pre-treatment attention, $\Psi$ reduces to the average treatment effect among attentive respondents, i.e., $\Psi=\E[Y_i(1)-Y_i(0)\mid W_i=1]$. 
In this case, the target subgroup is the same under both treatment conditions.

When engagement represents post-treatment manipulation, $\Psi$ instead represents the average treatment effect among respondents who would be successfully manipulated if assigned to treatment, regardless of their engagement status under control.
In the literature on manipulation checks, some researchers have studied $\Psi$ using a partial identification approach \citep{hazlett2026post}.
However, $\Psi$ is different from another popular causal quantity of interest in the literature, which is the average effect among ``always-manipulated'' respondents \citep{aronow2019note, tyler2025bounding}, defined as,
\begin{align*}
    \mathbb{E}[Y_i(1) - Y_i(0) \mid W_i(1) = W_i(0) = 1].
\end{align*}

This quantity targets respondents who would be manipulated regardless of treatment assignment, but this subgroup is often small and may not even exist in some experiments where the control condition is no manipulation (rather than comparing two different manipulations).
In addition, membership in this subgroup is never directly observed \citep{frangakis2002principal}. 
As a consequence, the identification of this quantity requires a stringent assumption, such as principal ignorability \citep{ding:lu:17}, or necessitates a partial identification approach \citep[e.g.,][]{aronow2019note, tyler2025bounding}.

\subsection{Identification when Engagement Status is Observed}
\label{sec:observed-identification}

For the sake of exposition, we first consider the case where the (true) engagement status $W_i$ is observed without measurement error.
Under Assumptions~\ref{ass:consistency}~and~\ref{ass:randomization}, the ATE is identified by the mean difference in outcome between the treatment and control groups with covariate adjustment:
\begin{align*}
    \tau := \E[Y_i(1)-Y_i(0)]
    = \E\bigl[\E[Y_i\mid T_i=1, \bX_i]-\E[Y_i\mid T_i=0, \bX_i]\bigr].
\end{align*}
We show that, under two additional assumptions, this ATE can be rescaled to identify the eATE.

First, we assume that treatment does not, on average, affect respondents who would not be engaged under either treatment condition. 
For pre-treatment attention, this requires inattentive respondents to be unaffected by treatment on average. 
For post-treatment manipulation, the same restriction applies to respondents who would not be successfully manipulated under either treatment condition.
We emphasize that this assumption needs to hold only on average and does not require the treatment effect to be zero for every respondent in this subgroup.
\begin{assumption}[Zero average treatment effect without engagement]
\label{ass:manipulation-effect}
Treatment has no average effect among respondents who
would not be engaged under either assignment:
\begin{align*}
    \E[Y_i(1)-Y_i(0)\mid W_i(1)=W_i(0)=0]=0.
\end{align*}
\end{assumption}

Second, we assume that treatment cannot reduce engagement for any respondent, thereby ruling out respondents who would be engaged under the control condition but not under the treatment. 
This monotonicity condition is naturally satisfied in many survey experiments. 
For pre-treatment attention checks, the assumption holds trivially because $W_i(1)=W_i(0)$. 
For post-treatment manipulation checks, the assumption is satisfied in one-sided treatment designs where manipulation is only active under treatment, ensuring that control respondents are never engaged, i.e., $W_i(0) = 0$ for every respondent. 
The assumption is also applicable in two-sided treatment designs where manipulation is active under both treatment and control, provided that the treatment condition is designed to be more engaging for every respondent than the control condition.
\begin{assumption}[Monotonicity]
\label{ass:monotonicity}
For every respondent,
\begin{align}
    W_i(1)\geq W_i(0).
    \label{eq:monotonicity}
\end{align}
\end{assumption}

Under Assumption~\ref{ass:monotonicity}, respondents with $W_i(1)=0$ are precisely those who would not be engaged under either treatment condition, i.e., $W_i(1)=W_i(0)=0$.
Assumption~\ref{ass:manipulation-effect} therefore implies that treatment has no average effect outside the target subgroup.
Applying the law of iterated expectations gives
\begin{align*}
    \E[Y_i(1)-Y_i(0)]
    &= \E[Y_i(1)-Y_i(0)\mid W_i(1)=1]
       \Pr(W_i(1)=1) \\
    &\quad+
       \E[Y_i(1)-Y_i(0)\mid W_i(1)=0]
       \Pr(W_i(1)=0) \\
    &= \Psi\,\Pr(W_i(1)=1).
\end{align*}
As consistency and randomization imply
$\Pr(W_i(1)=1)=\E[\Pr(W_i=1\mid T_i=1, \bX_i)]$, we can identify the eATE as,
\begin{align}
    \Psi
    = \frac{\E\bigl[\E[Y_i\mid T_i=1, \bX_i]-\E[Y_i\mid T_i=0, \bX_i]\bigr]}
           {\E[\Pr(W_i=1\mid T_i=1, \bX_i)]},
    \label{eq:manipulation-ratio}
\end{align}
provided that some treated respondents are engaged, i.e., $\Pr(W_i=1\mid T_i=1, \bX_i = \bx)>0$ (this assumption is formally stated below).
We note that in this setting, the ATE never exceeds the eATE in magnitude since the denominator is bounded between 0 and 1. This happens because we assume no effect without engagement (Assumption~\ref{ass:manipulation-effect}).

\subsection{Recovering Latent Engagement Status from Multiple Checks}
\label{sec:recovery}


The preceding identification result was obtained under the assumption that the true engagement status is directly observed. 
However, as discussed in Section~\ref{sec:motivation}, this assumption is too stringent in practice because attention and manipulation checks are frequently subject to measurement error.
Fortunately, we show that it is possible to recover the proportion of truly engaged respondents from multiple noisy attention or manipulation checks.
The use of multiple checks is a common setup in applied research and is recommended as a best practice \citep{berinsky2014separating}. 
We require three checks in order to identify the eATE. 

Specifically, for each respondent $i$, let $\widetilde W_{ji}$ denote the $j$th observed check, where $j\in \{1,2,3\}$. 
These checks may be binary, ordinal, or continuous.
Our goal is to recover the proportion of respondents who would be engaged under treatment, $\Pr(W_i(1)=1)$, which appears in the denominator of Equation~\eqref{eq:manipulation-ratio}.

\begin{figure}[t]
\centering
\begin{tikzpicture}[
  >=Latex,
  measured/.style={circle,draw,minimum size=1.05cm,inner sep=2pt},
  latent/.style={measured,dashed},
  bridge/.style={preaction={draw=white,line width=2.8pt,-}}
]
  \node[measured] (t)  at (-3, 0)   {$T$};
  \node[latent]   (w)  at ( 0, 0)   {$W$};
  \node[measured] (w1) at ( 3, 2.2) {$\widetilde W_1$};
  \node[measured] (w2) at ( 3, 0)   {$\widetilde W_2$};
  \node[measured] (w3) at ( 3,-2.2) {$\widetilde W_3$};
  \node[measured] (y)  at ( 5.7, 0) {$Y$};
  \node[measured] (x)  at ( 8, 2.6) {$\bX$};
  \draw[->] (w) to[bend left=8] (w1);
  \draw[->] (w) -- (w2);
  \draw[->] (w) to[bend right=8] (w3);
  \draw[->] (w1) to[bend left=8] (y);
  \draw[->] (w2) -- (y);
  \draw[->] (w3) to[bend right=8] (y);
  \draw[->] (w.-21) .. controls (2.65,-1.0) and (3.25,-1.19) .. (y.206);
  \draw[->] (t) -- (w);
  \draw[->] (t) to[bend left=45] (w1);
  \draw[->,bridge] (t) to[bend left=35] (w2);
  \draw[->] (t) to[bend right=25] (w3);
  \draw[->] (t.south)
    .. controls (-2.4,-4.0) and (5.1,-4.0) .. (y.south);
  \draw[->] (x) -- (w1);
  \draw[->,bridge] (x) -- (w2);
  \draw[->] (x) -- (y);
  \draw[->,bridge] (x.south)
    .. controls (8,-1.3) and (5.7,-2.6) .. (w3.east);
  \draw[->,bridge] (x.north west)
    .. controls (5.3,4.0) and (0.3,3.3) .. (w.north);
  \draw[->] (x.north)
    .. controls (7.7,4.9) and (-3,4.7) .. (t.north);
  \pgfresetboundingbox
  \path[use as bounding box] (-3.55,-3.2) rectangle (8.55,4.25);
\end{tikzpicture}
\caption{Directed Acyclic Graph for the proposed methodology. The dashed
circle denotes the unobserved true engagement status. We assume that,
conditional on $W$, $\bX$, and $T=1$, the three checks are mutually
independent. The three checks ($\widetilde W_1$, $\widetilde W_2$, and
$\widetilde W_3$) and the engagement status $W$ can be pre-treatment,
in which case there are no direct arrows from $T$ to these variables.}
\label{fig:measurement-dag}
\end{figure}
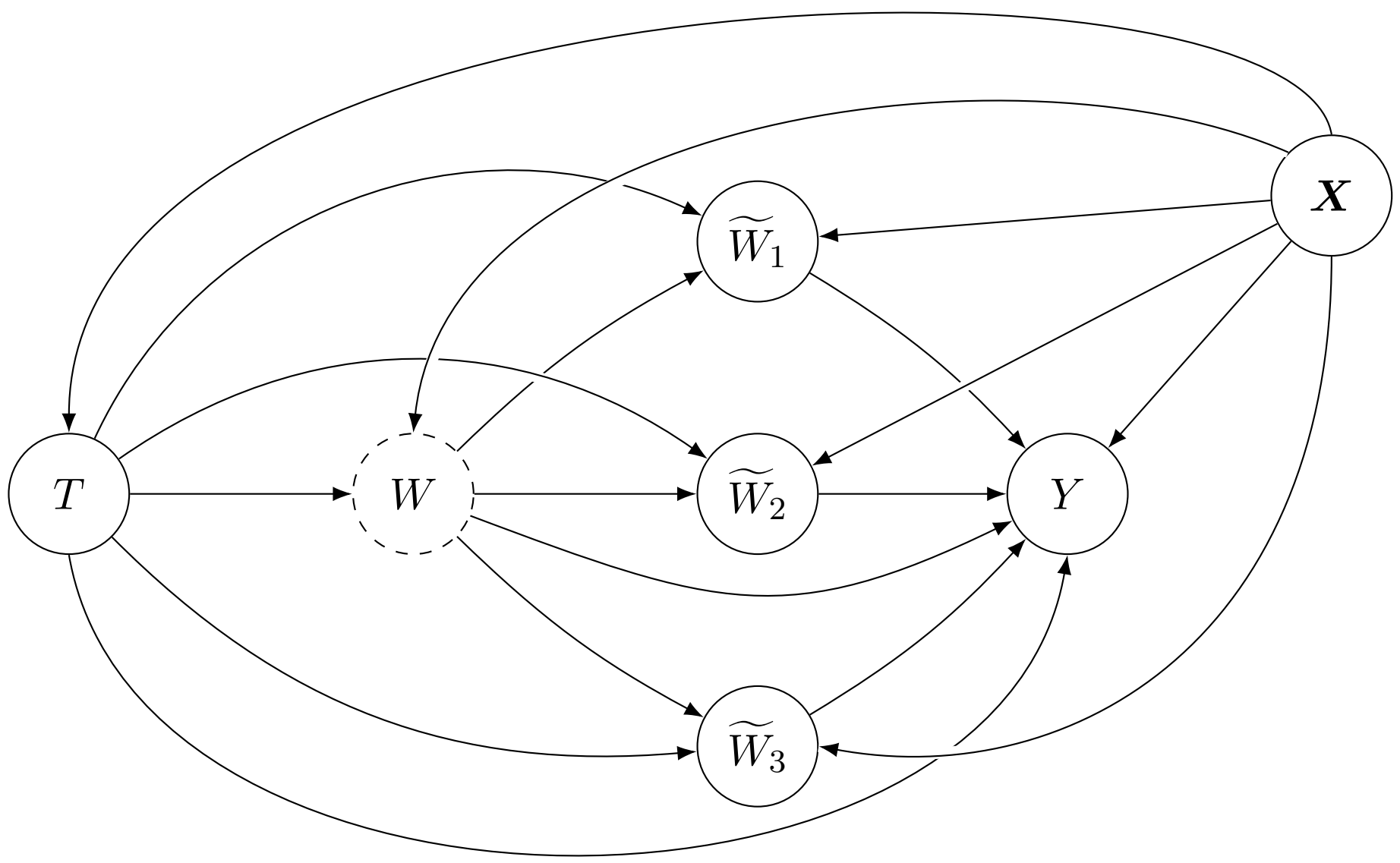

We impose two assumptions regarding the quality of checks. 
First, we assume that, among the treated respondents, the measurement errors of the checks are mutually independent given true engagement status and pre-treatment characteristics.
This assumption is formalized as follows:
\begin{assumption}[Conditional independence of measurement errors among checks]
\label{ass:measurement-independence}
\begin{align}
 \widetilde W_{1i} \ \indep \ \widetilde W_{2i}
 \ \indep \ \widetilde W_{3i} & \mid W_i,T_i=1,\bX_i.
 \label{eq:measurement-independence}
\end{align}
Or equivalently, $\P(\widetilde W_{1i} , \widetilde W_{2i} , \widetilde W_{3i} \mid W_i, T_i=1, \bX_i) = \P(\widetilde W_{1i} \mid W_i, T_i=1, \bX_i) \P(\widetilde W_{2i} \mid W_i, T_i=1, \bX_i) \P(\widetilde W_{3i} \mid W_i, T_i=1, \bX_i)$.
\end{assumption}
Figure~\ref{fig:measurement-dag} graphically illustrates the requirement of Assumption~\ref{ass:measurement-independence}. 
Among the treated respondents, we allow the checks to be associated with one another, but only through the true latent engagement status and the observed pre-treatment covariates. 
For example, a common comprehension process that
affects two or more checks beyond $W_i$ and $\bX_i$ would violate the required assumption.

Second, we require that, among treated respondents, the true latent engagement variable can take both values within each covariate stratum, and checks can distinguish engaged from unengaged respondents on average.
This condition is analogous to the relevance condition in the instrumental variable approach. 
Specifically, among treated respondents with the same pre-treatment covariates,
engaged respondents must have a higher average score on each check than unengaged ones, meaning that the engaged respondents are more likely to pass than unengaged respondents. 
\begin{assumption}[Positivity and Relevance]
\label{ass:recovery}
For almost every $\bx$, the following conditions hold:
\begin{enumerate}[label=(\roman*),leftmargin=2em]

\item (Positivity) $0 < \Pr(W_i = 1 \mid T_i = 1, \bX_i = \bx) < 1$.

 \item (Relevance of checks) For $j\in \{1,2,3\}$,
 \[
  \E[\widetilde W_{ji}\mid W_i=1,T_i=1,\bX_i=\bx]
    >
  \E[\widetilde W_{ji}\mid W_i=0,T_i=1,\bX_i=\bx].
 \]
\end{enumerate}
\end{assumption}
Under Assumption~\ref{ass:measurement-independence}, these conditions have observable implications, which we leverage to develop a statistical test of whether they hold in a given application (see Section~\ref{sec:testing}).

Under these assumptions, we can identify the conditional probability of engagement among treated respondents. 
\begin{proposition}[Recovery of latent engagement probabilities]
\label{prop:recovery}
Suppose that Assumptions~\ref{ass:measurement-independence}~and~\ref{ass:recovery} hold.
Then, for almost every $\bx$, the conditional engagement
probability is identified as
\begin{align}
    \Pr(W_i=1\mid T_i=1,\bX_i=\bx)
    &=
    1 - \bm e_0^\top \bm G_1(1,\bx)^{-1}
    \E[\bm V_{1i}\mid T_i=1,\bX_i=\bx],
    \label{eq:engagement-recovery}
\end{align}
where $\bm e_0=(1,0)^\top$, $\bm V_{1i}:=(1,\widetilde W_{1i})^\top$, and
$\bm G_1(1,\bx)$ is the matrix of conditional mean vectors
defined in Equation~\eqref{eq:check-mean-matrix} below and identifiable from the observed data.
If, in addition, Assumptions~\ref{ass:consistency}--\ref{ass:monotonicity} hold, then we have,
\begin{align}
    \Psi
    &=
    \frac{\E\bigl[\E[Y_i\mid T_i=1,\bX_i]-\E[Y_i\mid T_i=0,\bX_i]\bigr]}
         {\E[\Pr(W_i=1\mid T_i=1,\bX_i)]}.
    \label{eq:latent-subgroup-ratio}
\end{align}
\end{proposition}
Appendix~\ref{app:recovery} provides the proof of the proposition and the recovery and calibration formulas used for estimation. 
This result builds on the growing literature on measurement error and proxy variables and their applications to social science research \citep[e.g.,][]{kuroki2014measurement, zhou2024causal, mccartan2025estimating, nakamura2025surrogate, chasalow2026probabilistic, egami2026debiased}.

The essence of the identification argument relies on the following factorization of the conditional moments of the checks used in the literature \citep[see, e.g.,][for similar ideas]{hu2008identification, zhou2024causal, nakamura2025surrogate, egami2026debiased, guo2026proximal}. 
For $j=1,2$, let $\bm V_{ji}:=(1,\widetilde W_{ji})^\top$, which augments the $j$th check with a constant.
By the law of iterated expectations and the conditional independence of the first two checks under Assumption~\ref{ass:measurement-independence}, we have
\begin{align}
    &\E[\bm V_{1i}\bm V_{2i}^\top
        \mid T_i=1,\bX_i=\bx]
    \nonumber\\
    &\quad=
    \sum_{w=0}^1
    \E[\bm V_{1i}\bm V_{2i}^\top
        \mid W_i=w,T_i=1,\bX_i=\bx]
    \Pr(W_i=w\mid T_i=1,\bX_i=\bx)
    \nonumber\\
    &\quad=
    \sum_{w=0}^1
    \E[\bm V_{1i}\mid W_i=w,T_i=1,\bX_i=\bx]
    \E[\bm V_{2i}\mid W_i=w,T_i=1,\bX_i=\bx]^\top
    \nonumber\\
    &\qquad\quad{}\times\Pr(W_i=w\mid T_i=1,\bX_i=\bx)
    \nonumber\\
    &\quad=
    \bm G_1(1,\bx)
    \begin{bmatrix}
        \Pr(W_i=0\mid T_i=1,\bX_i=\bx) & 0\\
        0 & \Pr(W_i=1\mid T_i=1,\bX_i=\bx)
    \end{bmatrix}
    \bm G_2(1,\bx)^\top,
    \label{eq:probability-factorization}
\end{align}
for almost every $\bx$, where, for $j=1,2$ and $t\in\{0,1\}$, we define
\begin{align}
    \bm G_j(t,\bx)
    &:=
    \begin{bmatrix}
        1 & 1\\
        \E[\widetilde W_{ji}\mid W_i=0,T_i=t,\bX_i=\bx]
        &
        \E[\widetilde W_{ji}\mid W_i=1,T_i=t,\bX_i=\bx]
    \end{bmatrix}. \label{eq:check-mean-matrix}
\end{align}
Here, the columns of $\bm G_j(t,\bx)$ contain the conditional
mean vectors of $\bm V_{ji}$ for unengaged and engaged
respondents, respectively. Thus, the observed moment
matrix is a weighted sum of the products of these
subgroup-specific mean vectors, with weights given by
the subgroup probabilities.

Assumption~\ref{ass:measurement-independence} also implies that the third check is jointly independent of the first two checks conditional on engagement status and covariates among treated respondents. 
The conditional moment weighted by $\widetilde{W}_{3i}$ therefore admits another factorization:
\begin{align}
    &\E[\widetilde{W}_{3i}\bm V_{1i}\bm V_{2i}^\top
        \mid T_i=1,\bX_i=\bx]
    \nonumber\\
    = \ & 
    \bm G_1(1,\bx)
    \begin{bmatrix}
        \Pr(W_i=0\mid T_i=1,\bX_i=\bx)\,m_1(0,\bx) & 0\\
        0 & \Pr(W_i=1\mid T_i=1,\bX_i=\bx)\,m_1(1,\bx)
    \end{bmatrix}
    \bm G_2(1,\bx)^\top,
    \label{eq:probability-factorization2}
\end{align}
where
$m_1(w,\bx):=\E[\widetilde{W}_{3i}\mid W_i=w,T_i=1,\bX_i=\bx]$.
The left-hand sides of Equations~\eqref{eq:probability-factorization}~and~\eqref{eq:probability-factorization2} are identified from the observed distribution, whereas their right-hand sides contain the same unknown subgroup probabilities and check means. 
The only difference is that the second factorization additionally weights each subgroup by its conditional mean of the third check.

This common structure allows us to separate the unknown quantities. Specifically, multiplying the second moment matrix on the right by the inverse of the first cancels the subgroup probabilities and $\bm G_2(1,\bx)$:
\begin{align*}
    &\E[\widetilde{W}_{3i}\bm V_{1i}\bm V_{2i}^\top
        \mid T_i=1,\bX_i=\bx] \
        \E[\bm V_{1i}\bm V_{2i}^\top
            \mid T_i=1,\bX_i=\bx]^{-1}\\
    = \ &
    \bm G_1(1,\bx)
    \begin{bmatrix}
        m_1(0,\bx) & 0\\
        0 & m_1(1,\bx)
    \end{bmatrix}
    \bm G_1(1,\bx)^{-1}.
\end{align*}
Here, the right-hand side is an eigenvalue decomposition (EVD) of a matrix identified from the observed distribution.
Although an eigendecomposition is not unique in general, three features of this problem make identification possible.
First, the relevance condition regarding the third check in Assumption~\ref{ass:recovery} ensures that the eigenvalues are distinct, identifying the eigenvectors up to scale and order. 
Second, because the first entry of each column of $\bm G_1(1,\bx)$ equals one, normalizing each eigenvector to have first entry one fixes its scale. 
Finally, the strict ordering of check means in Assumption~\ref{ass:recovery} then assigns the eigenvector with the larger second entry to engaged respondents. 
This identifies the matrix of check means $\bm G_1(1,\bx)$, and this is sufficient to identify the conditional engagement probability $\Pr(W_i = 1 \mid T_i = 1, \bX_i = \bx)$ since
\begin{align*}
    \E[ \bm V_{1i}
        \mid T_i=1,\bX_i=\bx] &= \sum_{w}  \E[ \bm V_{1i}
        \mid W_i = w, T_i=1,\bX_i=\bx] \Pr(W_i = w \mid T_i = 1, \bX_i = \bx)\\
        &= \bm G_1(1,\bx) 
        \begin{bmatrix}
        \Pr(W_i = 0 \mid T_i = 1, \bX_i = \bx)\\
        \Pr(W_i = 1 \mid T_i = 1, \bX_i = \bx)
        \end{bmatrix}.
\end{align*}

\subsection{Connection with Instrumental Variable Approach}

We now discuss how the proposed approach is related to instrumental variables methods, which are often used to address the failure of manipulation \citep{mathur2025statistical_m}.
Consider a setting in which $W_i(t)=1$ represents successful manipulation under treatment assignment $T_i = t$ for $t=0,1$.
Under monotonicity, i.e., $W_i(1) \geq W_i(0)$, which is equivalent to Assumption~\ref{ass:monotonicity}, and the exclusion restriction (i.e., $T_i$ affects $Y_i$ only through $W_i$), the complier average treatment effect can be identified as \citep{angr:imbe:95} 
\begin{align*}
    &\E[Y_i(W_i =1) - Y_i(W_i =0)
    \mid W_i(1) = 1, W_i(0) = 0]\\
   = &
    \frac{\E\bigl[\E[Y_i \mid T_i = 1,\bX_i] - \E[Y_i \mid T_i = 0,\bX_i]\bigr]}
    {\E\bigl[\Pr(W_i = 1 \mid T_i = 1,\bX_i)
    - \Pr(W_i = 1 \mid T_i = 0,\bX_i)\bigr]}.
\end{align*}

This expression can also be applied using a proxy measure, so long as the proxy preserves the difference in mean engagement between the treatment and control conditions, i.e., 
$$
\E\bigl[\E[\widetilde W_{1i} \mid T_i=1,\bX_i]- \E[\widetilde W_{1i} \mid T_i=0,\bX_i]\bigr]
= \E\bigl[\E[W_i \mid T_i=1,\bX_i] - \E[W_i \mid T_i=0,\bX_i]\bigr].
$$ 
Importantly, under a one-sided treatment design (i.e., $\Pr(W_i(0) = 1) = 0$), both approaches target the same principal stratum.


This comparison highlights two key differences between the instrumental variable approach and our proposed methodology. 
First, our approach does not require the exclusion restriction, allowing treatment assignment to directly affect the outcome even when engagement status remains unchanged. 
In particular, we allow nonzero average treatment effect among respondents who would be engaged under both treatment conditions:
\begin{align*}
    \E[Y_i(1)-Y_i(0)
    \mid W_i(1)=1, W_i(0)=1] \neq 0.
\end{align*}
In contrast, the instrumental variable approach assumes that this average treatment effect is zero, which may be too restrictive in practice.
For example, when treatment is designed to induce anxiety, some respondents may remain anxious regardless of treatment assignment.
If the treatment affects the outcomes of these respondents, then the exclusion restriction will be violated. 

Second, the instrumental variable approach ignores potential measurement error in manipulation checks. 
In contrast, our approach allows for biased proxies and does not place a restriction on the conditional distribution of each check given engagement status, treatment assignment, and pre-treatment covariates, so long as the conditional independence (Assumption~\ref{ass:measurement-independence}) and relevance (Assumption~\ref{ass:recovery}) conditions are satisfied.

\section{Estimation and Inference}
\label{sec:estimation}

In this section, building on the identification results above, we develop an estimation procedure that accommodates flexible covariate modeling and provides valid uncertainty quantification.
To develop an estimator of $\Psi$ and characterize its uncertainty, there are two challenges we must overcome.
First, the identification argument recovers the measurement model separately at each covariate value $\bX_i = \bx$. 
When covariates are continuous or define many small strata, separately estimating the required conditional moments can be unstable. 
Second, inference must account for estimation of the latent engagement probabilities $\Pr(W_i = 1 \mid T_i = 1, \bX_i = \bx)$ and measurement functions $m_1(w,\bx)$.

We address these challenges by approximating the dependence on covariates using spline functions and then propagating first-stage estimation error using methods from semiparametric statistics. 
Formally, let $q(\bx):=\Pr(W_i=1\mid T_i=1,\bX_i=\bx)$ denote the conditional engagement probability and $Q:= \Pr(W_i(1)=1) = \E[q(\bX_i)]$.
Together with the four check means in $\bm G_1(1,\bx)$ and $\bm G_2(1,\bx)$ and the two third-check means $m_1(0,\bx)$ and $m_1(1,\bx)$, there are seven unknown scalar functions of the covariates.

Let $L_n$ denote the number of basis functions, which may increase
with the sample size $n$, and let
$\varphi_1(\bx),\ldots,\varphi_{L_n}(\bx)$ be a user-specified
spline basis satisfying $\varphi_\ell(\bx)\geq0$ and
$\sum_{\ell=1}^{L_n}\varphi_\ell(\bx)=1$.
Researchers may choose among different spline bases; in the
empirical applications of Section~\ref{sec:applications},
we use piecewise-linear (degree-one) spline bases.
Specifically, we approximate the seven unknown scalar
functions by
\begin{align}
    q(\bx;\bm\theta)
    &=\sum_{\ell=1}^{L_n}\varphi_\ell(\bx)\omega_{1\ell},
    \ \
    \bm G_j(1,\bx;\bm\theta)
    =
    \begin{bmatrix}
        1 & 1\\[1mm]
        \displaystyle\sum_{\ell=1}^{L_n}
            \varphi_\ell(\bx)\lambda_{j0\ell}
        &
        \displaystyle\sum_{\ell=1}^{L_n}
            \varphi_\ell(\bx)\lambda_{j1\ell}
    \end{bmatrix},
    \ \
    m_1(w,\bx;\bm\theta)
    =\sum_{\ell=1}^{L_n}\varphi_\ell(\bx)\xi_{w\ell},
    \label{eq:spline-model}
\end{align}
for $j\in\{1,2\}$ and $w\in\{0,1\}$, and $\bm\theta$ is the parameter vector that collects $\{\omega_{1\ell},\lambda_{j0\ell}, \lambda_{j1\ell},\xi_{0\ell},\xi_{1\ell}\}_{\ell=1}^{L_n}$ for $j=1,2$.
We impose $0<\omega_{1\ell}<1$, $\lambda_{j0\ell}<\lambda_{j1\ell}$, and $\xi_{0\ell}<\xi_{1\ell}$ for every $\ell$ and $j$.
Because the basis functions are nonnegative and sum to one, these restrictions ensure positivity and proxy relevance in Assumption~\ref{ass:recovery}.
We use a common basis for convenience, but each function has its own coefficients, allowing different relationships with the covariates. 
However, using a common basis is not required, and different bases may be used for different functions.

We estimate $\bm\theta$ using the conditional moment restrictions in Equations~\eqref{eq:probability-factorization}~and~\eqref{eq:probability-factorization2}. 
Let $\bm\rho_i(\bm\theta)$ be the eight-dimensional vector obtained by stacking the columns of the residual matrices, defined as,
\begin{align*}
\bm\rho_i(\bm\theta)
:=
\begin{bmatrix}
    \operatorname{vec}\{\bm R_{1i}(\bm\theta)\}\\[1mm]
    \operatorname{vec}\{\bm R_{2i}(\bm\theta)\}
\end{bmatrix}
\in\mathbb R^8,
\end{align*}
where $\operatorname{vec}(\cdot)$ stacks the columns of a matrix, and
\begin{align}
    &\bm R_{1i}(\bm\theta) := \bm V_{1i}\bm V_{2i}^\top
    -\bm G_1(1,\bX_i;\bm\theta)
    \begin{bmatrix}
        1-q(\bX_i;\bm\theta) & 0\\
        0 & q(\bX_i;\bm\theta)
    \end{bmatrix}
    \bm G_2(1,\bX_i;\bm\theta)^\top,
    \nonumber\\[1mm]
    &\bm R_{2i}(\bm\theta) := \widetilde{W}_{3i}\bm V_{1i}\bm V_{2i}^\top
    \nonumber\\
    & \qquad\qquad-
    \bm G_1(1,\bX_i;\bm\theta)
    \begin{bmatrix}
        \{1-q(\bX_i;\bm\theta)\}m_1(0,\bX_i;\bm\theta) & 0\\
        0 & q(\bX_i;\bm\theta)m_1(1,\bX_i;\bm\theta)
    \end{bmatrix}
    \bm G_2(1,\bX_i;\bm\theta)^\top.
    \label{eq:matrix-residual}
\end{align}
When the candidate functions equal the population
functions, the identification restrictions imply
\begin{align}
    \E[\bm\rho_i(\bm\theta)\mid T_i=1,\bX_i]=\bm0.
    \label{eq:conditional-moment}
\end{align}
We therefore estimate the coefficients by minimizing a
regularized sum of squared moment residuals among treated
respondents:
\begin{align}
    \widehat{\bm\theta}
    \in\argmin_{\bm\theta\in\Theta}
    \left\{
    \frac1n\sum_{i=1}^n
    \mathbbm{1}\{T_i = 1\}
    \|\bm\rho_i(\bm\theta)\|_2^2
    +
    \lambda_n \Omega_n(\bm\theta)
    \right\},
    \label{eq:minimum-distance}
\end{align}
where $\Theta$ is the parameter space, $\Omega_n$ is the regularization term of the nuisance functions, and $\lambda_n \geq 0$ controls the strength of regularization. 
The regularization term is especially important when researchers include a large number of pre-treatment covariates.  
Throughout the paper, we use the squared $L_2$ norm of adjacent coefficient differences used in \cite{chen2012estimation}:
\[
\Omega_n = \sum_{r=1}^{R-1}
\frac{\norm{\bm\theta_{r+1}-\bm\theta_r}_2^2}
     {\tau_{r+1}-\tau_r},
\]
where $\tau_1<\cdots<\tau_R$ are the $R$ knots, including the boundary knots, and $\bm\theta_r$ collects the nuisance-function coefficients at the $r$th knot.
For a piecewise-linear basis, $R=L_n$, and it is recommended to choose $L_n$ of order $n^{1/5}$ and $\lambda_n$ of order $L_n/n$ \citep{chen2012estimation}.
However, researchers may use other forms of regularization depending on the application.

This objective is nonlinear due to the cross products of the spline basis and can be computationally challenging.
However, we can simplify the computation by projecting the residual $\bm\rho_i(\bm\theta)$ onto another sufficiently rich basis of $\bX_i$ \citep{chen2009efficient,chen2012estimation}. 
Specifically, we consider a separate spline basis $\bm h(\bx)$ that spans the products of up to four original spline basis functions in Equation~\eqref{eq:spline-model} (i.e., three check means and an engagement probability).
Because $\bm\rho_i(\bm\theta)$ contains products of at most four candidate functions, this basis represents every such moment exactly. 

By the orthogonality of least squares, we can then write the residuals as the sum of projection and projection error:
\begin{align*}
 \frac{1}{n}\sum_{i=1}^n \mathbbm{1}\{T_i = 1\}
 \|\bm\rho_i(\bm\theta)\|_2^2
 &=
 \frac{1}{n}\sum_{i=1}^n \sum_{r = 1}^8 T_i
 \biggl\| 
  \bm h(\bX_i)^\top\widehat{\bm\beta}_{r}(\bm\theta)
 \biggr\|_2^2 +
 \frac{1}{n}\sum_{i=1}^n \sum_{r = 1}^8  T_i
 \biggl\|\bm\rho_{ir}(\bm\theta)
   - \bm h(\bX_i)^\top\widehat{\bm\beta}_{r}(\bm\theta) \biggr \|_2^2,
\end{align*}
where $\rho_{ir}(\bm\theta)$ is the $r$-th component of the residual $\bm\rho_i(\bm\theta)$ ($r \in \{1, \cdots, 8\}$) and $\widehat{\bm\beta}_{r}$ is its projection onto the column space spanned by $\bm h(\bX_i)$ defined as 
\begin{align*}
    \widehat{\bm\beta}_{r}(\bm\theta)
 &=
 \left\{
   \sum_{i=1}^n \mathbbm{1}\{T_i = 1\}
   \bm h(\bX_i)\bm h(\bX_i)^\top
 \right\}^{-1}
 \sum_{i=1}^n \mathbbm{1}\{T_i = 1\}
 \bm h(\bX_i)\bm\rho_{ir}(\bm\theta).
\end{align*}
Since the basis $\bm h(\bx)$ contains all the terms in the original spline basis, the projection error does not depend on $\bm\theta$ and thus Equation~\eqref{eq:minimum-distance} can be simplified as
\begin{align}
    \widehat{\bm\theta}
    \in\argmin_{\bm\theta\in\Theta}
     \frac1n\sum_{r=1}^{8} \biggl(
 \widehat{\bm\beta}_{r}(\bm\theta)^\top
 \left\{
   \sum_{i=1}^n \mathbbm{1}\{T_i = 1\} 
   \bm h(\bX_i)\bm h(\bX_i)^\top
 \right\}
 \widehat{\bm\beta}_{r}(\bm\theta) \biggr) + \lambda_n \Omega_n(\bm\theta)  .\label{eq:minimum-distance-simplified} 
\end{align}
The optimization problem in Equation~\eqref{eq:minimum-distance-simplified} is computationally easier since the inner term (Gram matrix) does not depend on the parameter $\bm\theta$ and thus we only need to compute once.

We then propagate the first-stage estimation error, building on recent developments in the semiparametric statistics literature (e.g., \citealt{chernozhukov2018double, tsiatis2006semiparametric}). 
Specifically, we consider the \emph{influence function}, a function that satisfies
\begin{align*}
    \E[\phi(T_i, Y_i, \bX_i, \widetilde W_{1i},\widetilde W_{2i},\widetilde W_{3i}; \bm\eta,  \Psi)] = 0
\end{align*}
and
\begin{align*}
    \sqrt n(\widehat\Psi-\Psi)
    =\frac1{\sqrt n}\sum_{i=1}^n\phi(T_i, Y_i, \bX_i, \widetilde W_{1i},\widetilde W_{2i},\widetilde W_{3i}; \bm\eta, \Psi)+o_p(1),
\end{align*}
where $\bm\eta$ is the collection of nuisance functions and
$o_p(1)$ refers to the error that is negligible as $n \to \infty$. For the influence function below, the first condition serves as a moment condition to estimate the parameter of interest $\Psi$, whereas the second expression is useful to derive its asymptotic variance using the central limit theorem.
The influence function is a key concept in the recent development of semiparametric statistics, since estimating equations based on it are insensitive to first-order errors in the nuisance estimates, a property known as \textit{Neyman orthogonality} \citep{chernozhukov2018double}. In our setup, the influence function is derived as follows.

\begin{proposition}[Influence function]
\label{prop:observed-if}
Under Assumptions~\ref{ass:consistency}--\ref{ass:recovery}, provided that the component
influence functions below are square-integrable, an influence function for
$\Psi$ is given by
\begin{align}
    &\phi(T_i, Y_i, \bX_i, \widetilde W_{1i},\widetilde W_{2i},\widetilde W_{3i}; \bm\theta, \mu_0, \mu_1, Q, \Psi)\nonumber \\
    &\qquad:=\frac{\phi_{1}(T_i, Y_i, \bX_i; \mu_1)-\phi_{0}(T_i, Y_i, \bX_i; \mu_0)-\Psi\phi^Q(T_i, \bX_i, \widetilde W_{1i},\widetilde W_{2i},\widetilde W_{3i}; \bm\theta)  }{Q},
    \label{eq:ate-if}
\end{align}
where $\phi_t(T_i,Y_i,\bX_i;\mu_t)$ is an uncentered influence function for $\Psi_t:=\E[\mu_t(\bX_i)]$ defined as
\begin{align*}
    \phi_t(T_i,Y_i,\bX_i;\mu_t)
    &:=
    \mu_t(\bX_i)
    +\frac{\mathbf1\{T_i=t\}}{\pi_t(\bX_i)}
     \{Y_i-\mu_t(\bX_i)\},
\end{align*}
with $\mu_t(\bx):=\E[Y_i\mid T_i=t,\bX_i=\bx]$.
Similarly,
$\phi^Q(T_i,\bX_i,\widetilde W_{1i},\widetilde W_{2i},
\widetilde W_{3i};\bm\theta)$ is an uncentered influence function for $Q=\Pr(W_i(1)=1)$ defined as
\begin{align*}
    \phi^Q(T_i, \bX_i, \widetilde W_{1i},\widetilde W_{2i},\widetilde W_{3i}; \bm\theta)
    &:=q(\bX_i)
      +\frac{\mathbf1\{T_i=1\}}{\pi_1(\bX_i)}
       \{1-S_{0i}+C_i-q(\bX_i)\},
\end{align*}
where
\begin{align*}
    S_{ji}
    &:=\frac{[\bm G_j(1,\bX_i)]_{2,2}
                   -\widetilde W_{ji}}
            {[\bm G_j(1,\bX_i)]_{2,2}
                   -[\bm G_j(1,\bX_i)]_{2,1}},
    \qquad j=1,2,
    &S_{0i}&:=S_{1i}S_{2i}.\\
    C_i
    &:=\frac{\widetilde W_{3i}-m_1(1,\bX_i)}
            {m_1(1,\bX_i)-m_1(0,\bX_i)}
       \{S_{1i}+S_{2i}-2S_{0i}\}.
\end{align*}

\end{proposition}
The proof is given in Appendix~\ref{app:observed-if}.
Here, we can see that the uncentered influence functions take the canonical form of augmented inverse propensity score except that $\phi^Q$ uses $1-S_{0i}+C_i$ instead of $W_i$ since it is latent and thus needs to be estimated.

We then obtain the proposed estimator by substituting fitted
functions into the estimating equations for
$\Psi_0$, $\Psi_1$, and $Q$. We use $K$-fold
cross-fitting so that each respondent's contribution
is evaluated using functions estimated without that
respondent's fold \citep{chernozhukov2018double, van2011targeted}. For a fixed $K\geq2$, the procedure
is as follows.
\begin{enumerate}[label=\arabic*.,leftmargin=2em]
    \item Randomly partition the observations into
    $K$ folds, $\mathcal I_1,\ldots,\mathcal I_K$,
    of approximately equal size.

    \item For each fold $k$, estimate the measurement model
using observations outside $\mathcal I_k$:
\begin{align*}
        \widehat{\bm\theta}^{(-k)}
    \in\argmin_{\bm\theta\in\Theta}
     \frac{1}{n - |\mathcal I_k|}\sum_{r=1}^{8} \biggl(
 \widehat{\bm\beta}_{r}(\bm\theta)^\top
 \left\{
   \sum_{i\notin\mathcal I_k} T_i 
   \bm h(\bX_i)\bm h(\bX_i)^\top
 \right\}
 \widehat{\bm\beta}_{r}(\bm\theta)\biggr)+ \lambda_n \Omega_n(\bm\theta) 
\end{align*}
Substitute $\widehat{\bm\theta}^{(-k)}$ into Equation~\eqref{eq:spline-model} to obtain the fitted engagement probabilities, the means of the first two checks, and subgroup-specific means of the third check.
Also estimate $\mu_t(\bx)=\E[Y_i\mid T_i=t,\bX_i=\bx]$ using the training observations in arm $t$, for $t=0,1$.

    \item Compute
    \begin{align}
        \widehat\Psi_t
        &=\frac1n\sum_{k=1}^K\sum_{i\in\mathcal I_k}
        \left[
            \hat\mu_t^{(-k)}(\bX_i)
            +\frac{\mathbf1\{T_i=t\}}{\pi_t(\bX_i)}
             \{Y_i-\hat\mu_t^{(-k)}(\bX_i)\}
        \right],
        \label{eq:overall-estimator}\\
        \widehat Q
        &=\frac1n\sum_{k=1}^K\sum_{i\in\mathcal I_k}
        \left[
            \hat q^{(-k)}(\bX_i)
            +\frac{\mathbf1\{T_i=1\}}{\pi_1(\bX_i)}
             \{1-\widehat S_{0i}^{(-k)}
                +\widehat C_i^{(-k)}
                -\hat q^{(-k)}(\bX_i)\}
        \right].
        \label{eq:probability-estimator}
    \end{align}
    and set
    \begin{align}
        \widehat\Psi
        =\frac{\widehat\Psi_1-\widehat\Psi_0}
               {\widehat Q}.
        \label{eq:ate-estimators}
    \end{align}
\end{enumerate}

Finally, we establish the asymptotic distribution of
the proposed estimator under the regularity conditions of Assumption~\ref{cond:inference} in Appendix~\ref{app:asymptotic}.

\begin{proposition}[Asymptotic normality]
\label{prop:asymptotic}
Under Assumptions~\ref{ass:consistency}--\ref{cond:inference},
\begin{align}
    \sqrt n(\widehat\Psi-\Psi)
   \ \xrightarrow{d}\ \mathcal N\bigl(0,\E[\phi(T_i, Y_i, \bX_i, \widetilde W_{1i},\widetilde W_{2i},\widetilde W_{3i}; \bm\theta, \mu_0, \mu_1, Q, \Psi)^2]\bigr).
    \label{eq:asymptotic-normality}
\end{align}
\end{proposition}
The proof is given in Appendix~\ref{app:asymptotic}.  We can thus calculate the standard errors by
\begin{align*}
    \widehat{\operatorname{se}}(\widehat\Psi)
    =\left\{
        \frac1{n^2}\sum_{k=1}^K\sum_{i\in\mathcal I_k}
        (\hat\phi_i-\bar{\hat\phi})^2
    \right\}^{1/2}.
\end{align*}
where $\hat\phi_i :=\phi\bigl(T_i, Y_i, \bX_i, \widetilde W_{1i},
                 \widetilde W_{2i},\widetilde W_{3i};\,\hat{\bm\theta}^{(-k)},
       \hat\mu_0^{(-k)},\hat\mu_1^{(-k)},
       \widehat Q,\widehat\Psi\bigr)$ is the estimated influence function and  $\bar{\hat\phi}=n^{-1}\sum_{k=1}^K\sum_{i\in\mathcal I_k} \hat\phi_i$ is its empirical average.
An asymptotic 95\% confidence interval is then obtained as
$\widehat\Psi\pm1.96\,
\widehat{\operatorname{se}}(\widehat\Psi)$.

\section{Testing the Relevance Assumption}
\label{sec:testing}

The key identification assumptions of the proposed methodology are the conditional independence of checks among treated individuals (Assumption~\ref{ass:measurement-independence}) and the relevance of checks (Assumption~\ref{ass:recovery}). 
Although these assumptions involve unobserved engagement status, the relevance condition has observable implications under the conditional independence assumption. 
Specifically, we can assess the relevance assumption through associations among observed variables. 
The following proposition provides the covariance identities underlying these diagnostics.

\begin{proposition}[Covariance identity]
\label{prop:covariance}
Consider a treatment condition $t$, under which the conditional
independence restrictions in
Assumption~\ref{ass:measurement-independence} hold.
Then, we have,
\begin{align}
    \Cov(\widetilde W_{ji},\widetilde W_{j^\prime i}
         \mid T_i=t,\bX_i=\bx)
    &=d_t^j(\bx)d_t^{j^\prime}(\bx)
      \V(W_i\mid T_i=t,\bX_i=\bx),
      \qquad 1\leq j<j^\prime\leq3,
    \label{eq:proxy-covariance}
\end{align}
where
\begin{align*}
    d_t^j(\bx)
    &:=\E[\widetilde W_{ji}\mid W_i=1,T_i=t,\bX_i=\bx]
       -\E[\widetilde W_{ji}\mid W_i=0,T_i=t,\bX_i=\bx],
    \quad j\in \{1,2,3\}.
\end{align*}
\end{proposition}
The proof is given in Appendix~\ref{app:covariance}.

Proposition~\ref{prop:covariance} shows that a zero covariance between checks implies that at least one of three factors is zero: the first check fails to distinguish engaged respondents from the unengaged ones, the second check fails to distinguish these groups, or engagement status does not vary within the covariate stratum. 
In the treated group, the last factor is given by,
\begin{align*}
    \V(W_i\mid T_i=1,\bX_i=\bx)
    =q(\bx)\{1-q(\bx)\},
\end{align*}
which is strictly positive under the positivity condition given in Assumption~\ref{ass:recovery}.
Therefore, under conditional independence and positivity, the covariance between any two checks is nonzero if and only if both checks have different conditional means across engagement groups.
Moreover, the positive ordering imposed by the relevance condition implies that all three pairwise conditional covariances are strictly positive.
These observable implications allow us to assess the strength of the relevance condition without first recovering the latent engagement status.

We can implement these diagnostics using pairwise regressions with covariate adjustment. 
Specifically, among treated respondents, we consider the following regression for each pair
$(j,k)\in\{(1,2),(1,3),(2,3)\}$:
\begin{align}
    \widetilde W_{ji}
    &= \bm p_0(\bX_i)^\top\bm\zeta_{jk}
     + \widetilde W_{ki}\bm p_1(\bX_i)^\top\bm\nu_{jk}
     + e_{jk,i},
    \label{eq:proxy-test-regression}
\end{align}
where $\bm p_0(\bX_i)$ and $\bm p_1(\bX_i)$ are user-specified covariate basis vectors that include an intercept, and $\bm\zeta_{jk}$ and $\bm\nu_{jk}$ are pair-specific coefficient vectors.
The first term adjusts for the covariates, while the second allows the association between checks to vary with the covariates.
For each pair, we conduct an $F$-test of $H_0:\bm\nu_{jk}=\bm0$ to assess conditional association between checks. We then use an intersection--union test
to assess whether all three pairs exhibit conditional
association. Specifically, we test
\[
H_0:\bm\nu_{jk}=\bm0
\text{ for at least one pair }(j,k)
\quad\text{against}\quad
H_1:\bm\nu_{jk}\neq\bm0
\text{ for all three pairs}.
\]
The resulting $p$-value is
\[
p_{\mathrm{IUT}}=\max\{p_{12},p_{13},p_{23}\}.
\]
We reject the null hypothesis at level $\alpha$ if all three pairwise $p$-values are below $\alpha$.

\section{Simulation Studies}\label{sec:simulation}

In this section, we conduct simulation studies to examine the performance of the proposed estimator under varying degrees of proxy quality, considering both attention and manipulation checks. 

\subsection{Setup}\label{sec:simulation-design}

For $n = 1000$
respondents, we independently draw a pre-treatment covariate
$\bX_i\sim\operatorname{Uniform}(-1,1)$ and treatment assignment
$T_i\sim\operatorname{Bernoulli}(1/2)$. We then generate potential
engagement status as
\[
W_i(1)=\mathbbm{1}\{U_i\le0.6+0.2\bX_i\},
\qquad
W_i(0)=\mathbbm{1}\{U_i\le\gamma(0.6+0.2\bX_i)\}.
\]
where $U_i \sim \mathrm{Uniform}(0,1)$. We set $\gamma=1$ for attention, so that engagement is unaffected
by treatment, and $\gamma=0.4$ for manipulation, so that treatment
can increase engagement. 

Throughout the simulation, we assume that realized engagement
$W_i=W_i(T_i)$ is unobserved. Instead, we observe three binary checks, generated as
\[
\widetilde W_{ji}
=
\mathbbm{1}\{a_j+0.3\bX_i+\kappa(2W_i-1)+Z_{ji}>0\},
\qquad j\in\{1,2,3\},
\]
where $Z_{ji} \sim \mathcal{N}(0,1)$ ($j \in \{1,2,3\}$) is the error in the observed checks and $(a_1,a_2,a_3)=(-0.5,0,0.5)$ allows the checks to differ in
difficulty. Here, $\kappa$ controls the quality of checks: a larger $\kappa$ corresponds to more accurate measurement of $W_{i}$.
We finally generate potential outcomes as
\[
Y_i(t)=\bX_i^2+0.5W_i(0)+tW_i(1)(1+\bX_i + \{Z_{1i} + Z_{2i} + Z_{3i}\} )+\varepsilon_i,
\qquad t\in\{0,1\},
\]
where $\varepsilon_i\sim\operatorname{Uniform}(-1,1)$ is an outcome error, and we observe $Y_i=Y_i(T_i)$.

Under this data-generating process, we consider attention and manipulation checks and vary the check quality $\kappa \in \{1.0, 1.5, 2.0, 2.5\}$. 
For each configuration, we generate $1000$ Monte Carlo replications to evaluate the performance of the estimator.

In each replication, we implement the proposed estimator using five-fold cross-fitting and piecewise-linear splines in $\bX_i$.
Following the penalized sieve estimation literature \citep{chen2012estimation}, we use $L_n=4$ spline basis functions and set the regularization weight to $\lambda_n=0.004$.
Within each training sample, we estimate the conditional engagement probability and measurement functions using treated respondents by minimizing the objective in Equation~\eqref{eq:minimum-distance-simplified}.

We also estimate the conditional outcome means separately by treatment arm using the same spline basis.
We then evaluate the fitted functions on the held-out observations and combine their contributions across folds to obtain the corrected engagement probability estimate $\widehat Q$ and the eATE estimate $\widehat\Psi=(\widehat\Psi_1-\widehat\Psi_0)/\widehat Q$, as defined in Equations~\eqref{eq:probability-estimator}~and~\eqref{eq:ate-estimators}, using the known treatment assignment probabilities $\pi_0(\bx)=\pi_1(\bx)=1/2$.

We compare the proposed estimator with three conventional approaches. 
All three estimators adjust for the pre-treatment covariate $\bX_i$ using the same spline basis as the proposed estimator. 
First, we consider the intention-to-treat (ITT) effect by regressing outcomes on treatment assignment and the spline basis functions of $\bX_i$.
Unlike the eATE, this estimator averages treatment effects over both engaged and unengaged respondents. 
Second, for attention checks, we consider a subgroup analysis that estimates the treatment effect among respondents who pass all three observed checks, i.e., $\widetilde W_{1i}=\widetilde W_{2i} =\widetilde W_{3i}=1$, using the same covariate-adjusted regression. 
This approach treats the observed checks as accurate measures of engagement and is therefore subject to bias from measurement error.
Third, for manipulation checks, we consider an instrumental variable (IV) estimator that uses randomized treatment assignment as an instrument for an indicator of passing all three manipulation checks, adjusting for $\bX_i$ in both stages. 
This approach is also sensitive to measurement error in the checks and additionally relies on the exclusion restriction. 
In contrast, our proposed estimator explicitly accounts for measurement error
and does not require the standard IV exclusion restriction.

\subsection{Results}

\begin{figure}[p]
    \centering
    \includegraphics[width=1.0\linewidth]{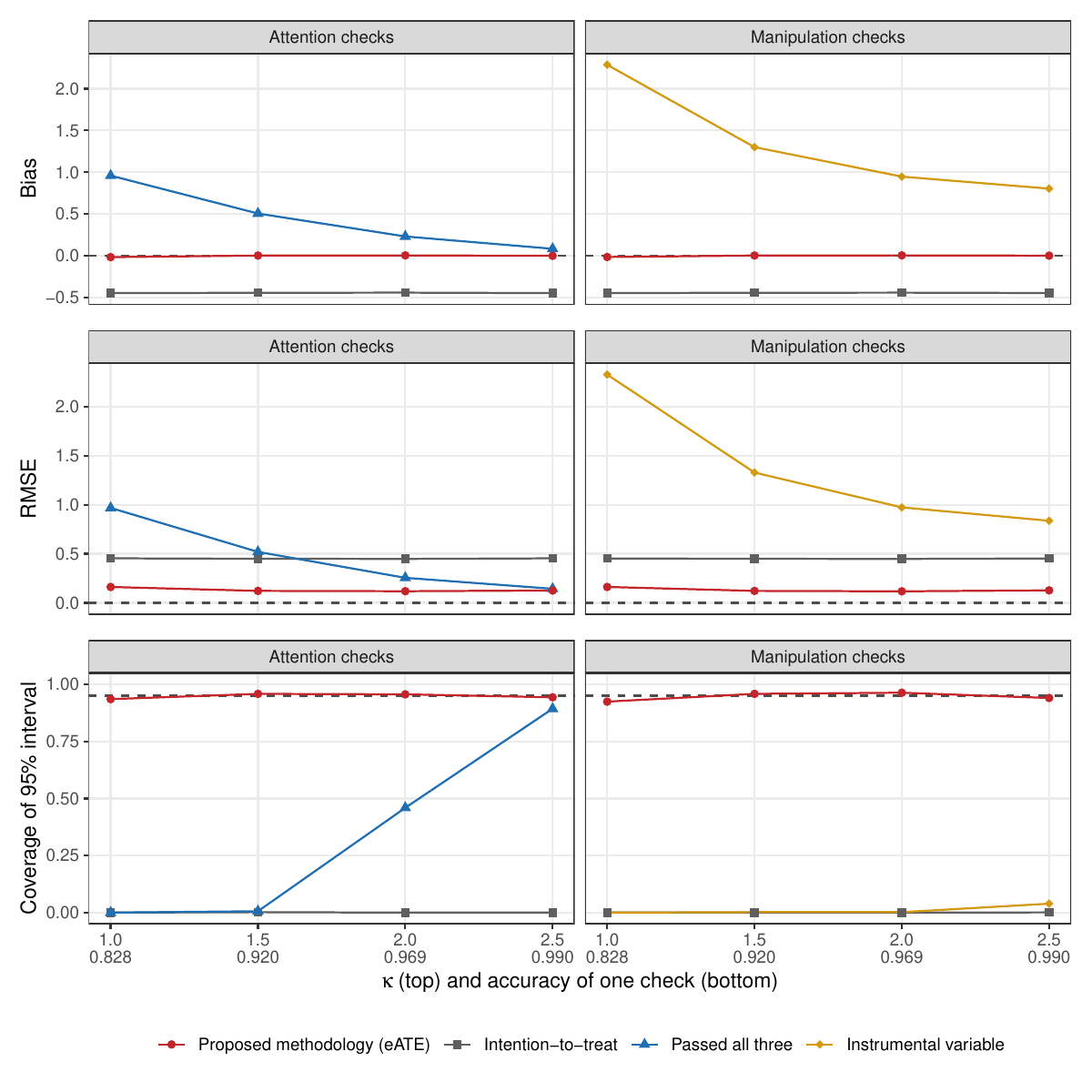}
    \caption{Simulation results comparing the proposed methodology with conventional estimators for attention checks (left) and manipulation checks (right). The three rows report bias,
root mean squared error (RMSE), and empirical coverage of
95\% confidence intervals for the engaged average treatment
effect (eATE). The horizontal axis shows the check-quality
parameter $\kappa$ (top) and the corresponding accuracy of
a single check (bottom). Each setting uses $n=1000$
respondents and $1000$ Monte Carlo replications.
The dashed horizontal lines indicate zero bias, zero RMSE,
and nominal 95\% coverage, respectively.
The conventional estimators include the intention-to-treat
estimator, the subgroup estimator restricted to respondents
who pass all three attention checks, and the instrumental
variable estimator for manipulation checks.}
    \label{fig_sim}
\end{figure}

Figure~\ref{fig_sim} presents the simulation results for attention checks in the left column and manipulation checks in the right column. 
Across all levels of check accuracy (horizontal axis), the proposed estimator (red solid circles) exhibits negligible bias (top row), low RMSE (middle), and empirical coverage (bottom) close to the nominal 95\% level.
In contrast, the conventional estimators generally exhibit substantial bias and poor coverage. 
In both attention and manipulation settings, the intention-to-treat estimator is biased toward zero because it averages treatment effects over both engaged and unengaged respondents. 
For pre-treatment attention checks, restricting the analysis to respondents who pass all three checks also produces a substantial bias when the checks are imperfect. 
Although this bias decreases as check accuracy increases, the subgroup estimator continues to exhibit undercoverage even when the accuracy of each check is relatively high. 
For post-treatment manipulation checks, the instrumental variable estimator exhibits substantial bias and poor coverage even when the checks are nearly perfectly accurate ($\kappa=2.5$). 
This bias persists because the exclusion restriction is violated, a problem that cannot be resolved by improving the accuracy of the observed checks alone.

\section{Results of Empirical Analyses}
\label{sec:applications}

In this section, we apply the proposed methodology to the two motivating examples introduced in Section~\ref{sec:motivation}: the framing experiment of \citet{berinsky2014separating}, which uses attention checks, and the digital media literacy experiment of \citet{guess2020digital}, which uses manipulation checks. 
In both applications, we compare our proposed approach with conventional methods to examine how accounting for measurement error in the checks affects the estimated treatment effects.

\subsection{Civil Liberties Framing Experiment}

We first apply the proposed methodology to the civil liberties framing experiment introduced in Section~\ref{example:attention}.
Specifically, we analyze the replication experiment of \citet{nelson1997media} administered in the second wave of Study 1 in \citet{berinsky2014separating}, a two-wave online panel conducted by Survey Sampling International in June--July 2011. 

Among 381 respondents with observed treatment assignment and outcome, five have missing attention check responses, leaving $N = 376$ respondents for the analysis sample.
As described in Section~\ref{example:attention}, our treatment of interest is assignment to the public safety frame rather than the free speech frame, which we code as treatment and control, respectively. 
The outcome is a binary indicator that is equal to one if the respondent thinks that the university should allow the KKK rally and zero otherwise.
We then construct four binary attention check variables indexed by their order of administration.
Specifically, $\widetilde W_{ji}$ equals one if respondent $i$ correctly answered the screener administered in position $j$, and zero otherwise, for $j=1,\ldots,4$.
The first screener was administered before the framing experiment, whereas the remaining three were administered afterwards.
Because screener types were randomly assigned to administration positions, the content of the screener corresponding to each $\widetilde W_{ji}$ can differ across respondents.

We then apply the proposed methodology to estimate the average treatment effect among engaged respondents or eATE. Specifically, we use the first three attention checks since their timing is the closest to the framing experiment we reanalyze.
We use five-fold cross-fitting and adjust for age and gender.
To model the dependence on these covariates, we use an
additive basis consisting of three piecewise-linear spline
basis functions for age and a gender indicator, and set
the regularization parameter to $\lambda_n=3/376\approx0.008$.
We also estimate the conditional outcome means separately by treatment condition using the same covariate basis. 

We compare the eATE estimate from our proposed methodology with the conventional estimates based on three different samples: all respondents in the analysis sample, respondents who passed the screener immediately preceding the experiment, and respondents who passed all four screeners. 
The first benchmark estimates the intention-to-treat (ITT) effect in the full analysis sample, while the other two restrict the sample based on observed screener responses. 
To ensure comparability, we adjust all three benchmark estimates for the same covariates used in our approach, with the same age spline specification and gender indicator.

\begin{figure}[t]
    \centering
    \includegraphics[width=1.0\linewidth]{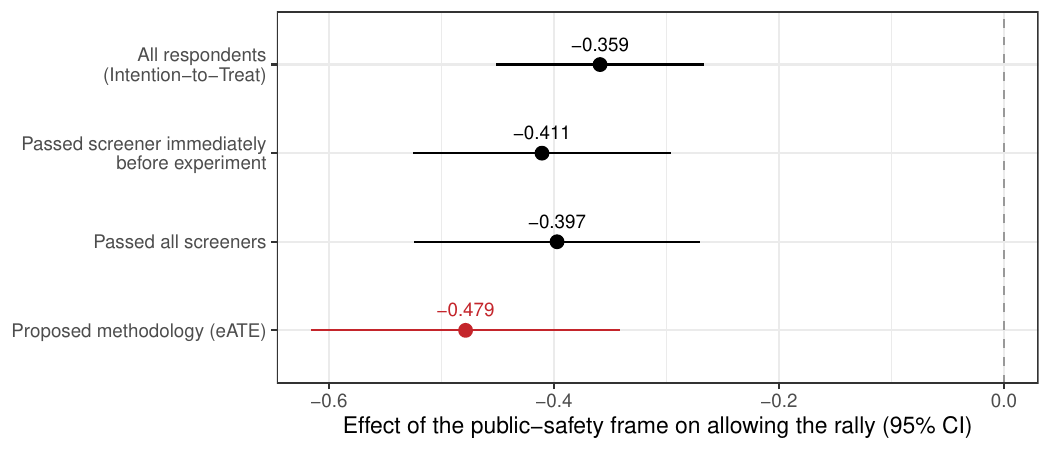}
    \caption{Estimated effects of the public safety frame
    relative to the free speech frame on support for allowing
    the KKK rally in the civil liberties framing experiment
    of \citet{berinsky2014separating}. The dashed vertical line indicates zero.
    The first estimator uses all respondents in the analysis
    sample. The second and third estimators restrict the
    analysis to respondents who passed the screener
    immediately preceding the experiment and those who
    passed all four screeners in the second wave, respectively.
    The proposed estimator accounts for measurement error
    in the first three screeners to estimate the average
    treatment effect among attentive respondents.}
    \label{fig:application1}
\end{figure}

Figure~\ref{fig:application1} reports the results.
Across all four approaches, we find that the public safety frame
substantially reduces respondents' probability of supporting the rally
relative to the free speech frame.
Specifically, the estimated average effect is $-0.359$
(95\% CI: $[-0.451,-0.267]$) among all respondents,
$-0.411$ (95\% CI: $[-0.525,-0.296]$) among respondents
who passed the screener immediately preceding the experiment,
and $-0.397$ (95\% CI: $[-0.524,-0.270]$) among
those who passed all four screeners.
Our proposed approach estimates the eATE of $-0.479$
(95\% CI: $[-0.616,-0.342]$) among engaged respondents,
corresponding to a 47.9 percentage-point reduction in support
for allowing the rally.
This estimate is approximately 12.0 percentage points larger
in magnitude than the ITT estimate among all respondents.
To assess the relevance condition required for identification, we conduct the pairwise $F$-tests described in Section~\ref{sec:testing} among respondents assigned to the public safety frame.
The resulting $F$-statistics are $7.9$ for checks 1 and 2,
$6.4$ for checks 1 and 3, and $10.6$ for checks 2 and 3, and the intersection--union test described in
Section~\ref{sec:testing} also rejects the null hypothesis
that at least one pair lacks conditional association
($p_{\mathrm{IUT}}<0.0001$). 
These results support the relevance condition.

These findings suggest that restricting the analysis to respondents who pass the observed attention checks may not fully account for measurement error in these checks.
In particular, even respondents who pass all four screeners may include inattentive respondents, while some attentive respondents may fail one or more screeners. 
By accounting for such measurement error under the identification assumptions, our proposed approach yields a larger treatment effect estimate in magnitude than the conventional subgroup analyses.
Although the confidence intervals overlap, the lower bound of the 95\% confidence interval shifts from $-0.451$ for the ITT estimate to $-0.616$ for the proposed eATE estimate, while the upper bound shifts from $-0.267$ to $-0.342$.
This pattern is consistent with conventional approaches understating the magnitude of the framing effect among attentive respondents.

\subsection{Digital Media Literacy Experiment}

We next apply the proposed methodology to the data from the digital media literacy experiment \citep{guess2020digital} introduced in Section~\ref{example:manipulation}. 
Specifically, we analyze the first wave of the U.S. YouGov survey, conducted from November 20 to December 27, 2018. 
The sample consists of 4,907 respondents, including 2,454 assigned to the digital media literacy intervention and 2,453 assigned to the control condition. 
In the experiment, each respondent was asked to evaluate four mainstream (true) and two false news headlines on a four-point scale ranging from 1 (not at all accurate) to 4 (very accurate).
We consider three outcomes: perceived accuracy of false news headlines, perceived accuracy of mainstream news headlines, and discernment, defined as the average difference in accuracy rating between mainstream and false news.
Following the original paper, we construct three binary manipulation checks indicating whether respondents correctly answered each comprehension question on their first attempt.
These questions assess their understanding of misleading web addresses, spelling errors, and corroboration by multiple trusted sources.

We then apply the proposed approach to estimate the average treatment effect among respondents who would be successfully manipulated under treatment (the eATE). 
Specifically, we use five-fold cross-fitting and adjust for multiple pre-treatment covariates, including age, gender, race, and education. 
For the first two outcomes, we conduct the analysis at the headline level, including headline fixed effects to account for differences across headlines. 
Because each respondent evaluates multiple headlines, we cluster standard errors at the respondent level and ensure that all observations from the same respondent belong to the same cross-fitting fold.
For discernment, we conduct the analysis at the respondent level. 
To model the dependence on covariates, we use an additive basis
consisting of five piecewise-linear spline basis functions for age
and indicators for gender, race, and education, and set
the regularization parameter to $\lambda_n=5/4907\approx0.0010$.

As benchmark comparisons, we replicate the two original estimators: the intention-to-treat (ITT) estimator and the instrumental variable (IV) estimator. 
The former estimates the effect of randomized assignment, whereas the latter uses treatment assignment as an instrument for observed treatment receipt, defined as passing all three comprehension questions on the first attempt among treated respondents. 
To ensure comparability with the proposed estimator, we also adjust for the same pre-treatment covariates (i.e., age, gender, race, and education) in all three analyses. 
Following the original analysis, we include headline fixed effects and cluster standard errors at the respondent level for the first two outcomes.

\begin{figure}[t]
    \centering
\includegraphics[width=1.0\linewidth]{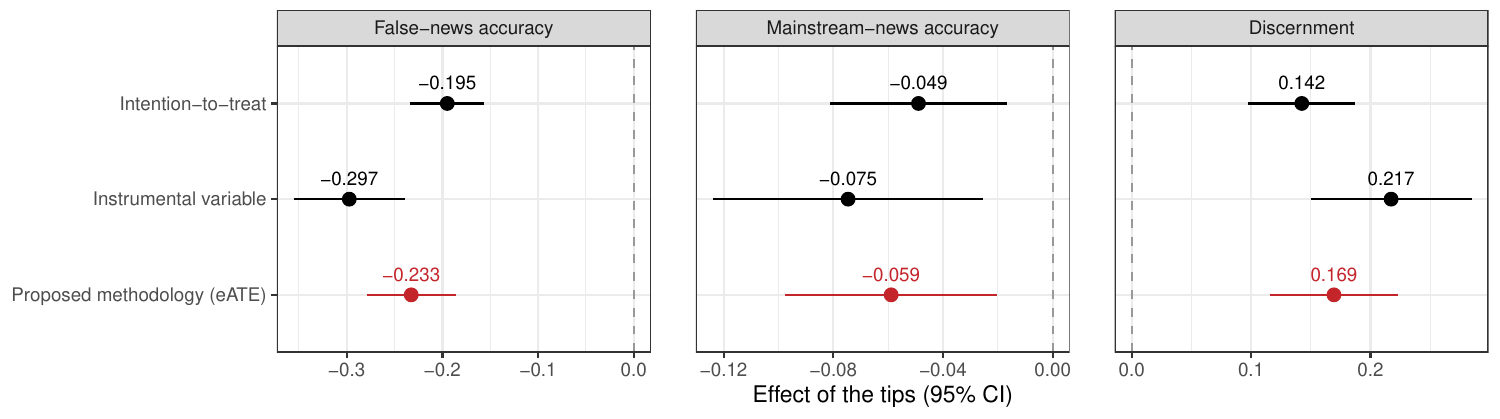}
    \caption{Estimated effects of the digital media literacy
intervention on perceived accuracy of false news (left),
perceived accuracy of mainstream news (middle), and
discernment (right) in \citet{guess2020digital}.
The dashed vertical line indicates zero. The intention-to-treat
estimator measures the effect of treatment assignment. The instrumental variable estimator
uses randomized treatment assignment as an instrument for passing the manipulation checks, which is defined as correctly answering all three
comprehension questions on the first attempt. The proposed
estimator of the eATE accounts for measurement error in these
checks to estimate the effect among respondents who would
be successfully manipulated under treatment.}
    \label{fig:application2}
\end{figure}

Figure~\ref{fig:application2} reports the results.
Across all three outcomes, we find that the intervention reduces the perceived accuracy of both false and mainstream news, with a larger reduction for false news, thereby improving respondents' ability to distinguish between mainstream and false news. 
Specifically, our proposed approach estimates that the
intervention reduces perceived accuracy by $0.233$ points for false news and $0.059$ points for mainstream news, and increases discernment by $0.169$ points.
To assess the relevance condition required for identification, we conduct the pairwise $F$-tests described in Section~\ref{sec:testing}. The resulting $F$-statistics are $21.8$, $16.9$, and $16.7$, with all three $p$-values below $0.0001$.
The intersection--union test also rejects the null hypothesis
that at least one pair lacks conditional association
($p_{\mathrm{IUT}}<0.0001$).
Overall, the estimates based on our proposed methodology are larger in absolute magnitude than the ITT
estimates but smaller than the IV estimates.

The difference between the proposed and IV methods is largely driven by the assumptions made about manipulation.
The IV approach assumes that manipulation is successful for those respondents who correctly answer all three comprehension questions on the first attempt. 
Under this definition, only $65.6\%$ of treated respondents are classified as successfully manipulated. 
In contrast, our proposed approach accounts for measurement
error in each manipulation check and estimates the proportion
of successfully manipulated respondents to be $0.836$
with a standard error of $0.015$, approximately 18
percentage points higher than the observed proportion.
Since both approaches rescale the intention-to-treat effect by the estimated proportion of successfully manipulated respondents, the smaller denominator used by the IV approach leads to larger treatment effect estimates.
These findings suggest that requiring respondents to pass all three checks for successful manipulation may be overly restrictive and may overstate the magnitude of the treatment effect among engaged respondents.

\section{Concluding Remarks}
\label{sec:conclusion}

In this paper, we propose a general statistical framework for analyzing survey experiments with attention and manipulation checks. 
Our unified approach accommodates both pre-treatment attention checks and post-treatment manipulation checks while accounting for measurement error in the observed checks. 
We show that it is possible to use three imperfect checks to recover the proportion of respondents who would be
engaged under treatment and identify the average treatment effect for this subgroup (eATE), under the assumption that these three checks are conditionally independent of one another given treatment, true engagement status, and the set of pre-treatment covariates. 
We then develop a flexible estimator, derive its asymptotic confidence intervals, and provide a diagnostic test for the key identification assumptions.
Through the reanalysis of a civil liberties framing experiment and a digital media literacy experiment, we demonstrate how accounting for measurement error can yield estimated treatment effects that are different from those based on conventional subgroup and instrumental variable analyses.

An important direction for future research, which we are currently pursuing, is to incorporate richer measurements of respondents' engagement. 
Eye tracking, for example, provides detailed information about which parts of a stimulus respondents view, how long they view them, and whether they return to particular pieces of information \citep[e.g.,][]{jenke2021using,jenke2025attention}. 
However, these measurements are also subject to measurement error. 
Future research could extend our framework to accommodate such richer, potentially high-dimensional measurements of respondents' engagement.

\bigskip
\bibliography{my}

\newpage
\appendix

\setcounter{equation}{0}
\setcounter{figure}{0}
\setcounter{table}{0}
\setcounter{section}{0}
\renewcommand {\theequation} {S\arabic{equation}}
\renewcommand {\thefigure} {S\arabic{figure}}
\renewcommand {\thetable} {S\arabic{table}}
\renewcommand {\thesection} {S\arabic{section}}

\section{Proofs}
\label{app:theory}

\subsection{Proof of Proposition~\ref{prop:recovery}}
\label{app:recovery}

We begin by writing Equations~\eqref{eq:probability-factorization}~and~\eqref{eq:probability-factorization2} as
\begin{align*}
 \E[\bm V_{1i}\bm V_{2i}^\top\mid T_i=1,\bX_i=\bx]
 &=\bm G_1(1,\bx)
   \operatorname{diag}\{1-q(\bx),q(\bx)\}
   \bm G_2(1,\bx)^\top,\\
 \E[\widetilde W_{3i}\bm V_{1i}\bm V_{2i}^\top\mid T_i=1,\bX_i=\bx]
 &=\bm G_1(1,\bx)
   \operatorname{diag}\{(1-q(\bx))m_1(0,\bx),
                         q(\bx)m_1(1,\bx)\}
   \bm G_2(1,\bx)^\top.
\end{align*}
The relevance assumption (Assumption~\ref{ass:recovery}(ii)) implies that $\bm G_j(1,\bx)$ is invertible for $j=1,2$ because their determinants are the strictly positive differences between the two check means.
The positivity condition (Assumption~\ref{ass:recovery}(i)) gives $0<q(\bx)<1$.
Thus, the first observed moment matrix is invertible, with
\begin{align*}
 &\E[\bm V_{1i}\bm V_{2i}^\top
           \mid T_i=1,\bX_i=\bx]^{-1}
           \ = \ \left\{\bm G_2(1,\bx)^\top\right\}^{-1}
   \operatorname{diag}\left\{\frac1{1-q(\bx)},\frac1{q(\bx)}\right\}
   \bm G_1(1,\bx)^{-1}.
\end{align*}
Multiplying the third-check-weighted matrix by this inverse yields
\begin{align*}
 &\E[\widetilde W_{3i}\bm V_{1i}\bm V_{2i}^\top\mid T_i=1,\bX_i=\bx]
  \E[\bm V_{1i}\bm V_{2i}^\top
              \mid T_i=1,\bX_i=\bx]^{-1}
 \nonumber\\
 &\qquad=\bm G_1(1,\bx)
   \operatorname{diag}\{m_1(0,\bx),m_1(1,\bx)\}
   \bm G_1(1,\bx)^{-1}.
\end{align*}
The relevance assumption about the third check ensures that the two eigenvalues are distinct. 
The right eigenvectors therefore identify the columns of $\bm G_1(1,\bx)$ up to scale and order.
Normalizing each eigenvector to have first entry one fixes its scale. 
The relevance condition assigns the column with the larger second entry to engaged respondents, identifying the
ordered columns and the corresponding third-check means.

Exchanging the roles of the checks gives
\begin{align*}
 &\E[\widetilde W_{3i}\bm V_{2i}\bm V_{1i}^\top\mid T_i=1,\bX_i=\bx]
  \E[\bm V_{2i}\bm V_{1i}^\top
              \mid T_i=1,\bX_i=\bx]^{-1}\\
 &\qquad=\bm G_2(1,\bx)
   \operatorname{diag}\{m_1(0,\bx),m_1(1,\bx)\}
   \bm G_2(1,\bx)^{-1}.
\end{align*}
Normalizing its eigenvectors and matching their eigenvalues to the labeled third-check means identifies $\bm G_2(1,\bx)$.
The recovered ordering agrees with the positive check-mean difference required for the second check.

Now, for either check, the following equality holds,
\begin{align*}
 \E[\bm V_{ji}\mid T_i=1,\bX_i=\bx]
 &=\bm G_j(1,\bx)
   \begin{bmatrix}1-q(\bx)\\q(\bx)\end{bmatrix}.
\end{align*}
Multiplication by the identified inverse of $\bm G_j(1,\bx)$ therefore identifies $q(\bx)$.

Define
\begin{align}
 \bm\ell_{j0}(1,\bx)
 &=\bm G_j(1,\bx)^{-\top}\bm e_0,
 \qquad j=1,2.
 \label{eq:calibration-coefficients}
\end{align}
Since these coefficients satisfy $\bm\ell_{j0}(1,\bx)^\top\bm G_j(1,\bx)=(1,0)$, we have
\begin{align*}
 \E[\bm\ell_{j0}(1,\bx)^\top\bm V_{ji}
       \mid W_i=w,T_i=1,\bX_i=\bx]
 &=1-w.
\end{align*}
In particular,
\begin{align*}
 1-q(\bx)
 &=\bm\ell_{j0}(1,\bx)^\top
   \E[\bm V_{ji}\mid T_i=1,\bX_i=\bx],
 \qquad j=1,2.
\end{align*}
The coefficients have the explicit form
\begin{align*}
 \bm\ell_{j0}(1,\bx)^\top
 &=\frac{\bigl([\bm G_j(1,\bx)]_{2,2},\ -1\bigr)}
        {[\bm G_j(1,\bx)]_{2,2}-[\bm G_j(1,\bx)]_{2,1}},
\end{align*}
where the 
denominator is strictly positive because of the relevance condition.

Consistency and randomization imply $q(\bx)=\Pr(W_i(1)=1\mid \bX_i=\bx)$.
For either $j=1,2$, the law of iterated expectations gives
\begin{align*}
 Q
 &=\int q(\bx)\,\d P_{\bX}(\bx)\nonumber\\
 &=1-\int\bm\ell_{j0}(1,\bx)^\top
       \E[\bm V_{ji}\mid T_i=1,\bX_i=\bx]\,\d P_{\bX}(\bx)
       \nonumber\\
 &=1-\E\bigl[\E[S_{ji}\mid T_i=1,\bX_i]\bigr].
\end{align*}
where $S_{ji}:=\bm\ell_{j0}(1,\bX_i)^\top\bm V_{ji}$.
Combining this result with Assumptions~\ref{ass:manipulation-effect}~and~\ref{ass:monotonicity}, as in Section~\ref{sec:observed-identification}, yields
\begin{align*}
 \Psi
 &=\frac{\E\bigl[\E[Y_i\mid T_i=1,\bX_i]-\E[Y_i\mid T_i=0,\bX_i]\bigr]}
 {1-\E\bigl[\E[S_{ji}\mid T_i=1,\bX_i]\bigr]},
 \qquad j=1,2.
\end{align*}
This proves the proposition.
\qed

\subsection{Proof of Proposition~\ref{prop:observed-if}}

Throughout the remainder of the appendix, we use the following notation
\begin{align*}
 \phi_{ti}
 &:=\phi_t(T_i,Y_i,\bX_i;\mu_t),\qquad t=0,1,\\
 \phi_i^Q
 &:=\phi^Q(T_i,\bX_i,\widetilde W_{1i},\widetilde W_{2i},
          \widetilde W_{3i};\bm\theta),\\
 \phi_i
 &:=\phi(T_i,Y_i,\bX_i,\widetilde W_{1i},\widetilde W_{2i},
        \widetilde W_{3i};\bm\theta,\mu_0,\mu_1,Q,\Psi).
\end{align*}
These functions are evaluated at the true nuisance functions and parameter values. 
We first establish the calibration properties of the surrogates and derive the influence function when $W_i$ is observed.
We then show that replacing $W_i$ with $1-S_{0i}+C_i$ preserves the score expectations required for an observed-data influence function.

\subsubsection{Lemmas and their Proofs}

We begin by proving two lemmas.
\begin{lemma}[Calibration and the correction term]
\label{lem:calibration}
Under Assumptions~\ref{ass:measurement-independence}~and~\ref{ass:recovery}, for $w=0,1$ and almost every $\bx$,
\begin{align*}
 \E[S_{ji}\mid W_i=w,T_i=1,\bX_i=\bx]
 &=1-w,\qquad j=1,2,\\
 \E[S_{0i}\mid W_i=w,T_i=1,\bX_i=\bx]
 &=1-w,\\
 \E[C_i\mid W_i=w,T_i=1,\bX_i=\bx]
 &=0.
\end{align*}
Consequently,
\begin{align*}
 \E[1-S_{0i}+C_i\mid W_i=w,T_i=1,\bX_i=\bx]&=w,\\
 \E[1-S_{0i}+C_i\mid T_i=1,\bX_i=\bx]&=q(\bx).
\end{align*}
\end{lemma}

\begin{proof}
By Equation~\eqref{eq:calibration-coefficients}, we have,
\begin{align*}
 \E[S_{ji}\mid W_i=w,T_i=1,\bX_i=\bx]
 &=\bm\ell_{j0}(1,\bx)^\top
   \E[\bm V_{ji}\mid W_i=w,T_i=1,\bX_i=\bx]\\
 &=1-w,
\end{align*}
for $j=1,2$.
Since $S_{0i}=S_{1i}S_{2i}$, conditional independence of the first
two checks gives,
\begin{align*}
 \E[S_{0i}\mid W_i=w,T_i=1,\bX_i=\bx]
 &=\E[S_{1i}\mid W_i=w,T_i=1,\bX_i=\bx]
   \E[S_{2i}\mid W_i=w,T_i=1,\bX_i=\bx]\nonumber\\
 &=(1-w)^2=1-w.
\end{align*}
The last equality follows because $W_i$ is binary. Therefore,
\begin{align*}
 \E[S_{1i}+S_{2i}-2S_{0i}\mid W_i=w,T_i=1,\bX_i=\bx]
 \ = \ (1-w)+(1-w)-2(1-w)=0.
\end{align*}

Recall that the correction term is
\begin{align*}
 C_i
 &=\frac{\widetilde W_{3i}-m_1(1,\bX_i)}
          {m_1(1,\bX_i)-m_1(0,\bX_i)}
       (S_{1i}+S_{2i}-2S_{0i}).
\end{align*}
The third check is independent of the first two checks conditional on
$\{W_i,T_i=1,\bX_i\}$. Hence,
\begin{align*}
 &\E[C_i\mid W_i=w,T_i=1,\bX_i=\bx]\nonumber\\
 = \ &
 \frac{m_1(w,\bx)-m_1(1,\bx)}{m_1(1,\bx)-m_1(0,\bx)}
 \E[S_{1i}+S_{2i}-2S_{0i}\mid W_i=w,T_i=1,\bX_i=\bx]
 =0.
\end{align*}
Combining these equalities and then averaging over $W_i$ yields
\begin{align*}
 \E[1-S_{0i}+C_i\mid W_i=w,T_i=1,\bX_i=\bx]
 &=1-(1-w)+0=w,\nonumber\\
 \E[1-S_{0i}+C_i\mid T_i=1,\bX_i=\bx]
 &=\E[W_i\mid T_i=1,\bX_i=\bx]=q(\bx).
\end{align*}
\end{proof}

\begin{lemma}[Full-data influence functions]
\label{lem:full-if}
Suppose that Assumptions~\ref{ass:consistency}--\ref{ass:monotonicity} hold,$0<\pi_1(\bx)<1$ for almost every $\bx$, $Q>0$, and $\E[Y_i^2]<\infty$.
Assume also that $\phi_{0i}$, $\phi_{1i}$, and $\phi_i^{Q,\mathrm{full}}$ below are square-integrable.
If the realized engagement status $W_i$ is observed, an uncentered influence function for $Q$ and an influence function for $\Psi$ are given by
\begin{align*}
 \phi_i^{Q,\mathrm{full}}
 &:=\phi^{Q,\mathrm{full}}(T_i,\bX_i,W_i;q)
   =q(\bX_i)+\frac{\mathbf1\{T_i=1\}}{\pi_1(\bX_i)}
      \{W_i-q(\bX_i)\}, \\
 \phi_i^{\mathrm{full}}
 &:=\phi^{\mathrm{full}}(T_i,Y_i,\bX_i,W_i;
                        q,\mu_0,\mu_1,Q,\Psi)
   =\frac{\phi_{1i}-\phi_{0i}
          -\Psi\phi_i^{Q,\mathrm{full}}}{Q},
\end{align*}
where $\phi_{ti}$ is the uncentered influence function for
$\Psi_t$ defined in Proposition~\ref{prop:observed-if}.
\end{lemma}

\begin{proof}
Consider any regular parametric submodel of the full-data model
through the true law. Denote its score by $s^F$ and the marginal score
for $\bX_i$ by $s_{\bX}$. Then,
\begin{align*}
 \E[s^F]=0,\qquad \E[s_{\bX}]=0,\qquad
 \E[s^F\mid \bX_i]=s_{\bX}.
\end{align*}

We first derive the influence function for $\Psi_t$.
Denote differentiation along the submodel at the
true law by a dot.
Then, differentiating the conditional mean gives
\begin{align*}
 \dot\mu_t(\bx)
 &=\E\!\left[
    Y_i\{s^F-\E[s^F\mid T_i=t,\bX_i=\bx]\}
    \mid T_i=t,\bX_i=\bx\right]\\
 &=\E[\{Y_i-\mu_t(\bx)\}s^F\mid T_i=t,\bX_i=\bx].
\end{align*}
Since $\Psi_t=\int\mu_t(\bx)\,\d P_{\bX}(\bx)$,
\begin{align*}
 \dot\Psi_t
 &=\int\dot\mu_t(\bx)\,\d P_{\bX}(\bx)+\E[\mu_t(\bX_i)s_{\bX}]\\
 &=\E\!\left[
    \frac{\mathbf1\{T_i=t\}}{\pi_t(\bX_i)}
    \{Y_i-\mu_t(\bX_i)\}s^F\right]
   +\E[\{\mu_t(\bX_i)-\Psi_t\}s^F]\\
&=\E\!\left[
    \biggl(\frac{\mathbf1\{T_i=t\}}{\pi_t(\bX_i)}
    \{Y_i-\mu_t(\bX_i)\} + \{\mu_t(\bX_i)-\Psi_t\}\biggr)s^F\right]
\end{align*}
Similarly, differentiating $q(\bx)=\P(W_i=1\mid T_i=1,\bX_i=\bx)$
gives
\begin{align*}
 \dot q(\bx)
 &=\E\!\left[
    W_i\{s^F-\E[s^F\mid T_i=1,\bX_i=\bx]\}
    \mid T_i=1,\bX_i=\bx\right]\\
 &=\E[\{W_i-q(\bx)\}s^F\mid T_i=1,\bX_i=\bx].
\end{align*}
Therefore,
\begin{align*}
 \dot Q
 &=\int\dot q(\bx)\,\d P_{\bX}(\bx)+\E[q(\bX_i)s_{\bX}]
 =\E\!\left[
    \biggl(\frac{\mathbf1\{T_i=1\}}{\pi_1(\bX_i)}
    \{W_i-q(\bX_i)\} + \{q(\bX_i)-Q\}\biggr)s^F\right].
\end{align*}
Finally, since $\Psi=(\Psi_1-\Psi_0)/Q$, by quotient rule,
\begin{align*}
 \dot\Psi
 &=\frac{(\dot\Psi_1-\dot\Psi_0)Q
           -(\Psi_1-\Psi_0)\dot Q}{Q^2}
           =\frac{\dot\Psi_1-\dot\Psi_0-\Psi\dot Q}{Q}
           =\E\!\left[
    \frac{\phi_{1i}-\phi_{0i}
           -\Psi\phi_i^{Q,\mathrm{full}}}{Q}s^F\right],
\end{align*}
which establishes the desired result.
\end{proof}

\subsubsection{Proof of Proposition~\ref{prop:observed-if}}
\label{app:observed-if}

We now relate the full-data and observed-data influence functions,
following the score argument in \citet{zhou2024causal}.

\begin{proof}
Let $s^F$ be the score of any regular parametric submodel of the
full-data model, as in Lemma~\ref{lem:full-if}. Marginalizing the
full-data likelihood over $W_i$ gives the observed-data score
\begin{align*}
 s^O=\E[s^F\mid O_i].
\end{align*}
Since $\phi_i^Q$ is measurable with respect to $O_i$,
\begin{align*}
 \E[\phi_i^Q s^O]
 &=\E[\phi_i^Q\E[s^F\mid O_i]]
  =\E[\phi_i^Q s^F].
\end{align*}
Thus, by Lemma~\ref{lem:full-if}, it suffices to show that
\begin{align}
 \E[(\phi_i^Q-\phi_i^{Q,\mathrm{full}})s^F]=0
 \label{eq:observed-score-condition}
\end{align}
for every such score, and that $\E[\phi_i^Q]=Q$.

Under Assumption~\ref{ass:measurement-independence}, the
conditional full-data density in the treated arm factors as
\begin{align*}
 &p(W_i\mid T_i=1,\bX_i)
   \prod_{j=1}^3
       p(\widetilde W_{ji}\mid W_i,T_i=1,\bX_i)
   p(Y_i\mid W_i,\widetilde W_{1i},\widetilde W_{2i},
                       \widetilde W_{3i},T_i=1,\bX_i).
\end{align*}
Differentiating the log likelihood therefore yields
\begin{align*}
 s^F
 &=s_{\bX}+\mathbf1\{T_i=1\}
          (s_W+s_1+s_2+s_3+s_Y)
       +\mathbf1\{T_i=0\}s_{\mathrm{ctr}},
\end{align*}
where $s_{\mathrm{ctr}}$ is the score of the conditional
full-data distribution in the control arm. There is no treatment
score because the randomization probabilities are fixed.
The conditional scores satisfy
\begin{align*}
 \E[s_W\mid T_i=1,\bX_i]&=0,\\
 \E[s_j\mid W_i,T_i=1,\bX_i]&=0,
       \qquad j=1,2,3,\\
 \E[s_Y\mid W_i,\widetilde W_{1i},\widetilde W_{2i},
                     \widetilde W_{3i},T_i=1,\bX_i]&=0.
\end{align*}

Now, the difference between the two expressions for $Q$ is
\begin{align*}
 \phi_i^Q-\phi_i^{Q,\mathrm{full}}
 &=\frac{\mathbf1\{T_i=1\}}{\pi_1(\bX_i)}
       \{1-S_{0i}+C_i-W_i\}.
\end{align*}
To verify Equation~\eqref{eq:observed-score-condition}, we
first calculate the conditional expectation of this difference
given each check and the latent engagement status.

For the third check, conditional independence and
Lemma~\ref{lem:calibration} give
\begin{align*}
 \E[S_{0i}\mid W_i=w,\widetilde W_{3i},T_i=1,\bX_i=\bx]
 &=1-w,\\
 \E[S_{1i}+S_{2i}-2S_{0i}
       \mid W_i=w,\widetilde W_{3i},T_i=1,\bX_i=\bx]
 &=0.
\end{align*}
The third-check factor in $C_i$ is fixed under this conditioning.
Consequently,
\begin{align}
 &\E[1-S_{0i}+C_i-W_i
       \mid W_i=w,\widetilde W_{3i},T_i=1,\bX_i=\bx]\nonumber\\
 &\qquad=1-(1-w)
       +\frac{\widetilde W_{3i}-m_1(1,\bx)}
              {m_1(1,\bx)-m_1(0,\bx)}\times0-w
 =0.
 \label{eq:replacement-outcome}
\end{align}

For either of the first two checks, let $j\in\{1,2\}$.
Conditional independence of the checks implies
\begin{align*}
 \E[S_{0i}\mid W_i=w,\widetilde W_{ji},T_i=1,\bX_i=\bx]
 &=(1-w)S_{ji},\\
 \E[S_{1i}+S_{2i}-2S_{0i}
       \mid W_i=w,\widetilde W_{ji},T_i=1,\bX_i=\bx]
 &=S_{ji}+(1-w)-2(1-w)S_{ji}.
\end{align*}
Since $W_i$ is binary,
\begin{align*}
 \frac{m_1(w,\bx)-m_1(1,\bx)}{m_1(1,\bx)-m_1(0,\bx)}
 &=w-1.
\end{align*}
Hence,
\begin{align*}
 &\E[C_i\mid W_i=w,\widetilde W_{ji},T_i=1,\bX_i=\bx]\\
 &\qquad=(w-1)
       \{S_{ji}+(1-w)-2(1-w)S_{ji}\}\\
 &\qquad=(w-1)(2w-1)S_{ji}+(w-1)(1-w)\\
 &\qquad=(1-w)(S_{ji}-1),
\end{align*}
where the last equality uses $w^2=w$. Therefore,
\begin{align}
 &\E[1-S_{0i}+C_i-W_i
       \mid W_i=w,\widetilde W_{ji},T_i=1,\bX_i=\bx]\nonumber\\
 &\qquad=1-(1-w)S_{ji}+(1-w)(S_{ji}-1)-w=0,
 \qquad j=1,2.
 \label{eq:replacement-proxies}
\end{align}

To apply these identities to the score, Equations~\eqref{eq:replacement-outcome}
and~\eqref{eq:replacement-proxies} imply
\begin{align*}
 \E[1-S_{0i}+C_i-W_i\mid W_i,\widetilde W_{ji},T_i=1,\bX_i]&=0,
       \qquad j=1,2,3,\\
 \E[1-S_{0i}+C_i-W_i\mid W_i,T_i=1,\bX_i]&=0,\\
 \E[1-S_{0i}+C_i-W_i\mid T_i=1,\bX_i]&=0.
\end{align*}
The last two equalities follow by iterated expectations.
Since $s_{\bX}$ is a function of $\bX_i$, $s_W$ is a function of
$(W_i,\bX_i)$ in the treated arm, and each $s_j$ is a function of
$(\widetilde W_{ji},W_i,\bX_i)$ in that arm,
\begin{align*}
 \E[(1-S_{0i}+C_i-W_i) s_{\bX}\mid T_i=1,\bX_i]
 &=s_{\bX}\E[(1-S_{0i}+C_i-W_i)\mid T_i=1,\bX_i]=0,\\
 \E[(1-S_{0i}+C_i-W_i) s_W\mid T_i=1,\bX_i]
 &=\E\!\left[
       s_W\E[(1-S_{0i}+C_i-W_i)\mid W_i,T_i=1,\bX_i]
       \mid T_i=1,\bX_i\right]=0,\\
 \E[(1-S_{0i}+C_i-W_i) s_j\mid T_i=1,\bX_i]
 &=\E\!\left[
       s_j\E[1-S_{0i}+C_i-W_i\mid W_i,\widetilde W_{ji},T_i=1,\bX_i]
       \mid T_i=1,\bX_i\right]=0
\end{align*}
for $j = 1,2,3$. For the outcome score, $1-S_{0i}+C_i-W_i$ is measurable given $W_i$, the three
checks, and $\bX_i$. Thus,
\begin{align*}
 &\E[(1-S_{0i}+C_i-W_i) s_Y\mid W_i,\widetilde W_{1i},\widetilde W_{2i},
                          \widetilde W_{3i},T_i=1,\bX_i]\\
 &\qquad=(1-S_{0i}+C_i-W_i)\E[s_Y\mid W_i,\widetilde W_{1i},\widetilde W_{2i},
                          \widetilde W_{3i},T_i=1,\bX_i]=0.
\end{align*}
The control-arm contribution is zero because
$\mathbf1\{T_i=1\}\mathbf1\{T_i=0\}=0$.
Combining these results gives
\begin{align*}
 \E[(\phi_i^Q-\phi_i^{Q,\mathrm{full}})s^F]
 &=\E\!\left[
       \frac{\mathbf1\{T_i=1\}}{\pi_1(\bX_i)}(1-S_{0i}+C_i-W_i) s^F\right]\\
 &=\int\E\!\left[
       (1-S_{0i}+C_i-W_i)(s_{\bX}+s_W+s_1+s_2+s_3+s_Y)
       \mid T_i=1,\bX_i=\bx\right]\,\d P_{\bX}(\bx)\\
 &=0.
\end{align*}

It follows that
\begin{align*}
 \dot Q =\E[\phi_i^{Q,\mathrm{full}}s^F]=\E[\phi_i^Q s^F]
  =\E[\phi_i^Q s^O]
  =\E[(\phi_i^Q-Q)s^O].
\end{align*}
Moreover, Lemma~\ref{lem:calibration} implies
\begin{align*}
 \E[\phi_i^Q]
 &=\E[q(\bX_i)]
   +\int\left\{
       \E[1-S_{0i}+C_i\mid T_i=1,\bX_i=\bx]-q(\bx)
      \right\}\,\d P_{\bX}(\bx)=Q.
\end{align*}
Thus, $\phi_i^Q-Q$ is an observed-data influence function for $Q$.

Finally, $\phi_{ti}$ is observed-data measurable, so
\begin{align*}
 \dot\Psi_t
 &=\E[\phi_{ti}s^F]
  =\E[\phi_{ti}\E[s^F\mid O_i]]
  =\E[\phi_{ti}s^O].
\end{align*}
Applying the quotient rule,
\begin{align*}
 \dot\Psi
 &=\frac{\dot\Psi_1-\dot\Psi_0-\Psi\dot Q}{Q}=\E\!\left[
     \frac{\phi_{1i}-\phi_{0i}-\Psi\phi_i^Q}{Q}s^O
     \right]
  =\E[\phi_i s^O].
\end{align*}
Also,
\begin{align*}
 \E[\phi_i]
 &=\frac{\E[\phi_{1i}]-\E[\phi_{0i}]
             -\Psi\E[\phi_i^Q]}{Q}\\
 &=\frac{\Psi_1-\Psi_0-\Psi Q}{Q}=0.
\end{align*}
Thus, these
identities establish the influence function in
Equation~\eqref{eq:ate-if}.
\end{proof}

\newpage
\subsection{Proof of Proposition~\ref{prop:asymptotic}}
\label{app:asymptotic}

\begin{assumption}[Regularity Conditions]
\label{cond:inference}
There exist fixed constants $c_1>0$ and
$c_2\in(0,1/2)$, and a deterministic sequence
$\delta_n>0$ with $\delta_n\to0$, such that the
following conditions hold. We suppress the superscript $(-k)$
for notational simplicity.

\begin{enumerate}[label=(\alph*),leftmargin=2em]

    \item (Boundedness)
    The outcome and numerically coded checks satisfy
    \begin{align*}
        \Pr\!\left(
            \max\{|Y_i|,|\widetilde W_{1i}|,
                       |\widetilde W_{2i}|,|\widetilde W_{3i}|\}
            \leq c_1
        \right)=1.
    \end{align*}
    All estimated nuisance functions are also uniformly
    bounded in absolute value by $c_1$ almost surely,
    componentwise for vector- and matrix-valued functions.

    \item (Positivity and Relevance)
    For almost every $\bx$,
    \begin{align*}
        c_2
        &\leq \pi_t(\bx)\leq 1-c_2,
        \qquad t=0,1,\\
        c_2
        &\leq q(\bx)\leq 1-c_2,\\
        \sigma_{\min}\{\bm G_j(1,\bx)\}
        &\geq c_2,
        \qquad j=1,2,\\
        [\bm G_j(1,\bx)]_{2,2}
        -[\bm G_j(1,\bx)]_{2,1}
        &\geq c_2,
        \qquad j=1,2,\\
        |m_1(1,\bx)-m_1(0,\bx)|
        &\geq c_2.
    \end{align*}
    Here, $\sigma_{\min}$ denotes the smallest singular
    value. The estimated functions $\widehat q$,
    $\widehat{\bm G}_j$, and $\widehat m_1$ satisfy
    the corresponding bounds almost surely.

        \item (Nuisance Function Estimation)
    With probability tending to one, the nuisance
    estimates satisfy
    \begin{align}
        \E\!\left[
            \left|
                \widehat\mu_t(\bX_i)-\mu_t(\bX_i)
            \right|^2
        \right]^{1/2}
        &\leq \delta_n,
        \qquad t=0,1,
        \nonumber\\
        \E\!\left[
            \left|
                \widehat q(\bX_i)-q(\bX_i)
            \right|^2
        \right]^{1/2}
        &\leq \delta_n,
        \nonumber\\
        \E\!\left[
            \sum_{r=0}^1
            \left|
                [\widehat{\bm\ell}_{j0}(1,\bX_i)]_r
                -[\bm\ell_{j0}(1,\bX_i)]_r
            \right|^2
        \right]^{1/2}
        &\leq n^{-1/4}\delta_n,
        \qquad j=1,2,
        \nonumber\\
        \E\!\left[
            \left|
                \widehat m_1(w,\bX_i)-m_1(w,\bX_i)
            \right|^2
        \right]^{1/2}
        &\leq n^{-1/4}\delta_n,
        \qquad w=0,1.
        \label{eq:nuisance-rates}
    \end{align}
    Here, $r=0,1$ indexes the two entries of
    $\bm\ell_{j0}$.

    \item (Nondegenerate Variance)
    The influence function in
    Equation~\eqref{eq:ate-if} satisfies
    \begin{align*}
        0<\E[\phi_i^2]<\infty.
    \end{align*}

\end{enumerate}
\end{assumption}

\begin{proof}
Let $n_k=|\mathcal I_k|$. For each fold $k$, we condition on
the observations outside $\mathcal I_k$ and on the random partition.
Throughout the proof, $c$ denotes a
finite positive constant that may change from line to line.
Also, we use the following fact in STEP2. For any nonnegative measurable function $g$, iterated expectations and
Assumption~\ref{cond:inference}(b) imply
\begin{align*}
 \E[g(\bX_i)\mid T_i=1]
 &=\frac{\E[\pi_1(\bX_i)g(\bX_i)]}{\Pr(T_i=1)}
 \leq\frac1{c_2}\E[g(\bX_i)],
\end{align*}
since $\Pr(T_i=1)=\E[\pi_1(\bX_i)]\geq c_2$.
We use this bound when transferring the marginal nuisance-rate
conditions to the treated arm.

\paragraph{STEP1: Linearizing numerator} We first consider the marginal outcome means. Subtracting the true
uncentered influence function from its fitted counterpart gives
\begin{align}
 H_{ti}
 &:=\widehat\mu_t(\bX_i)
   +\frac{\mathbf1\{T_i=t\}}{\pi_t(\bX_i)}
      \{Y_i-\widehat\mu_t(\bX_i)\}-\phi_{ti}
 \nonumber\\
 &=\widehat\mu_t(\bX_i)-\mu_t(\bX_i)
   -\frac{\mathbf1\{T_i=t\}}{\pi_t(\bX_i)}
      \{\widehat\mu_t(\bX_i)-\mu_t(\bX_i)\}
 \nonumber\\
 &=\left(1-\frac{\mathbf1\{T_i=t\}}{\pi_t(\bX_i)}\right)
      \{\widehat\mu_t(\bX_i)-\mu_t(\bX_i)\},
 \qquad t=0,1.
 \label{eq:score-differences}
\end{align}
Consequently, Equation~\eqref{eq:overall-estimator} implies
\begin{align}
 \widehat\Psi_t-\Psi_t
 &=\frac1n\sum_{k=1}^K\sum_{i\in\mathcal I_k}
       \{\phi_{ti}+H_{ti}^{(-k)}\}-\Psi_t
 =\frac1n\sum_{i=1}^n(\phi_{ti}-\Psi_t)
   +\sum_{k=1}^K\frac{n_k}{n}
       \frac1{n_k}\sum_{i\in\mathcal I_k}H_{ti}^{(-k)}.
 \label{eq:component-decomposition}
\end{align}
For a fixed fold, iterated expectations give
\begin{align*}
 \E[H_{ti}]
 &=\E\!\left[
    \E\!\left[
     \left(1-\frac{\mathbf1\{T_i=t\}}{\pi_t(\bX_i)}\right)
     \{\widehat\mu_t(\bX_i)-\mu_t(\bX_i)\}
       \,\bigg|\,\bX_i\right]\right]\\
 &=\E\!\left[
     \{\widehat\mu_t(\bX_i)-\mu_t(\bX_i)\}
     \left\{1-\frac{\Pr(T_i=t\mid \bX_i)}{\pi_t(\bX_i)}\right\}
    \right]\\
 &=\E\!\left[
     \{\widehat\mu_t(\bX_i)-\mu_t(\bX_i)\}
     \left(1-\frac{\pi_t(\bX_i)}{\pi_t(\bX_i)}\right)\right]\\
 &=0.
\end{align*}
The third equality follows from randomization. The same conditioning
gives the second-moment bound
\begin{align*}
 \E[H_{ti}^2]
 &=\E\!\left[
     |\widehat\mu_t(\bX_i)-\mu_t(\bX_i)|^2
     \E\!\left[
       \left(1-\frac{\mathbf1\{T_i=t\}}{\pi_t(\bX_i)}\right)^2
       \,\bigg|\,\bX_i\right]\right]\\
 &=\E\!\left[
     |\widehat\mu_t(\bX_i)-\mu_t(\bX_i)|^2
     \{\pi_t(\bX_i)(1-\pi_t(\bX_i)^{-1})^2+(1-\pi_t(\bX_i))\}\right]\\
 &=\E\!\left[
      \frac{1-\pi_t(\bX_i)}{\pi_t(\bX_i)}
      |\widehat\mu_t(\bX_i)-\mu_t(\bX_i)|^2\right]\\
 &\leq\frac1{c_2}
      \E[|\widehat\mu_t(\bX_i)-\mu_t(\bX_i)|^2]
 \leq\frac1{c_2}\delta_n^2.
\end{align*}
The inequalities use Assumption~\ref{cond:inference}(b)--(c).
Conditional independence of the evaluation observations now implies
\begin{align*}
 \E\!\left[
    \left\{\frac{\sqrt n}{n_k}
        \sum_{i\in\mathcal I_k}H_{ti}\right\}^2\right]=\frac{n}{n_k^2}\sum_{i\in\mathcal I_k}\E[H_{ti}^2]
 =\frac{n}{n_k}\E[H_{ti}^2]\leq\frac{n}{n_kc_2}\delta_n^2.
\end{align*}
Here, the cross terms vanish because each $H_{ti}$ has conditional
mean zero. Thus, for every $\varepsilon>0$, conditional Chebyshev's
inequality yields
\begin{align*}
 \Pr\!\left(
    \left|\frac{\sqrt n}{n_k}\sum_{i\in\mathcal I_k}H_{ti}\right|
      >\varepsilon
      \,\bigg|\,\text{training observations and partition}\right)
 &\leq\frac{n}{n_k\varepsilon^2}\E[H_{ti}^2]
 \leq\frac{n}{n_k\varepsilon^2c_2}\delta_n^2.
\end{align*}
The right-hand side tends to zero because $n/n_k$ is bounded and
$\delta_n\to0$. The rate event also has probability tending to one,
so the fold average is $o_p(n^{-1/2})$ unconditionally. Since $K$ is
fixed, its weighted average over folds is $o_p(n^{-1/2})$. Therefore,
\begin{align}
 \widehat\Psi_t-\Psi_t
 &=\frac1n\sum_{i=1}^n(\phi_{ti}-\Psi_t)+o_p(n^{-1/2}),
 \qquad t=0,1.
 \label{eq:mean-expansion}
\end{align}

\paragraph{STEP2: Linearizing denominator} Next, we consider the engagement probability. For a fixed fold,
define the difference between the fitted and true uncentered
influence functions by
\begin{align*}
 H_i^Q
 &:=\widehat q(\bX_i)
     +\frac{T_i}{\pi_1(\bX_i)}
       \{1-\widehat S_{0i}+\widehat C_i-\widehat q(\bX_i)\}
       -\phi_i^Q\\
 &=\widehat q(\bX_i)-q(\bX_i)
     +\frac{T_i}{\pi_1(\bX_i)}
       \{-\widehat S_{0i}+S_{0i}+\widehat C_i-C_i
                          -\widehat q(\bX_i)+q(\bX_i)\}\\
 &=\left(1-\frac{T_i}{\pi_1(\bX_i)}\right)
       \{\widehat q(\bX_i)-q(\bX_i)\}
     +\frac{T_i}{\pi_1(\bX_i)}
       \{-\widehat S_{0i}+S_{0i}+\widehat C_i-C_i\}.
\end{align*}
Equation~\eqref{eq:probability-estimator} then gives
\begin{align}
 \widehat Q-Q
 &=\frac1n\sum_{k=1}^K\sum_{i\in\mathcal I_k}
       \{\phi_i^Q+H_i^{Q,(-k)}\}-Q
 \nonumber\\
 &=\frac1n\sum_{i=1}^n(\phi_i^Q-Q)
   +\sum_{k=1}^K\frac{n_k}{n}
       \frac1{n_k}\sum_{i\in\mathcal I_k}H_i^{Q,(-k)}
 \nonumber\\
 &=\frac1n\sum_{i=1}^n(\phi_i^Q-Q)
   +\sum_{k=1}^K\frac{n_k}{n}
       \left\{\frac1{n_k}\sum_{i\in\mathcal I_k}H_i^{Q,(-k)}
                  -\E[H_i^{Q,(-k)}]\right\}
+\sum_{k=1}^K\frac{n_k}{n}\E[H_i^{Q,(-k)}].
 \label{eq:probability-remainder-decomposition}
\end{align}
We bound the centered average and the population
expectation separately.

\paragraph{STEP2-1: Bounding surrogate engagement $S_{0i}$} To bound the centered average, we first control the error in the
measurement correction. Bounded checks imply
$|m_1(w,\bx)|\leq c_1$ and
$|[\bm G_j(1,\bx)]_{2,w+1}|\leq c_1$ for $w=0,1$.
Thus, the formula for $S_{ji}$ gives
\begin{align*}
 |S_{ji}|
 &\leq
  \frac{|[\bm G_j(1,\bX_i)]_{2,2}|+|\widetilde W_{ji}|}
       {|[\bm G_j(1,\bX_i)]_{2,2}-[\bm G_j(1,\bX_i)]_{2,1}|}\leq\frac{2c_1}{c_2},\qquad j=1,2.
\end{align*}
The fitted functions satisfy the same boundedness and relevance
conditions, so $|\widehat S_{ji}|\leq2c_1/c_2$ as well.
By Equation~\eqref{eq:calibration-coefficients},
\begin{align*}
 \widehat S_{ji}-S_{ji}
 &=\{\widehat{\bm\ell}_{j0}(1,\bX_i)
                   -\bm\ell_{j0}(1,\bX_i)\}^\top\bm V_{ji}=\sum_{r=0}^1
    \{[\widehat{\bm\ell}_{j0}(1,\bX_i)]_r
                   -[\bm\ell_{j0}(1,\bX_i)]_r\}[\bm V_{ji}]_r.
\end{align*}
Thus, we have
\begin{align*}
 |\widehat S_{ji}-S_{ji}|^2
 &\leq(1+|\widetilde W_{ji}|^2)
     \sum_{r=0}^1
       |[\widehat{\bm\ell}_{j0}(1,\bX_i)]_r
                    -[\bm\ell_{j0}(1,\bX_i)]_r|^2\leq(1+c_1^2)
     \sum_{r=0}^1
       |[\widehat{\bm\ell}_{j0}(1,\bX_i)]_r
                    -[\bm\ell_{j0}(1,\bX_i)]_r|^2.
\end{align*}
where the first inequality follows from the Cauchy--Schwarz inequality and the second from Assumption~\ref{cond:inference}(a).
Therefore, by taking conditional expectations, we obtain
\begin{align}
 \E[|\widehat S_{ji}-S_{ji}|^2\mid T_i=1]^{1/2}
 &\leq\sqrt{1+c_1^2}
   \E\!\left[\sum_{r=0}^1
       |[\widehat{\bm\ell}_{j0}(1,\bX_i)]_r
          -[\bm\ell_{j0}(1,\bX_i)]_r|^2\,\bigg|\,T_i=1\right]^{1/2}
 \nonumber\\
 &\leq\sqrt{\frac{1+c_1^2}{c_2}}
   \E\!\left[\sum_{r=0}^1
       |[\widehat{\bm\ell}_{j0}(1,\bX_i)]_r
          -[\bm\ell_{j0}(1,\bX_i)]_r|^2\right]^{1/2}
 \nonumber\\
 &\leq\sqrt{\frac{1+c_1^2}{c_2}}\,n^{-1/4}\delta_n,
 \qquad j=1,2.
 \label{eq:surrogate-ms-bound}
\end{align}
where the second inequality uses the covariate-moment bound at the beginning
of the proof (by taking $g(\bX_i) = \sum_{r=0}^1
       |[\widehat{\bm\ell}_{j0}(1,\bX_i)]_r
          -[\bm\ell_{j0}(1,\bX_i)]_r|^2$), and the last inequality uses Assumption~\ref{cond:inference}(c).

For the product surrogate, adding and subtracting
$S_{1i}\widehat S_{2i}$ gives
\begin{align*}
 \widehat S_{0i}-S_{0i} =\widehat S_{1i}\widehat S_{2i}-S_{1i}S_{2i} =\widehat S_{1i}\widehat S_{2i}-S_{1i}\widehat S_{2i}
       +S_{1i}\widehat S_{2i}-S_{1i}S_{2i}=\widehat S_{2i}(\widehat S_{1i}-S_{1i})
       +S_{1i}(\widehat S_{2i}-S_{2i}).
\end{align*}
The triangle inequality for the square root of the second moment
therefore implies
\begin{align}
 \E[|\widehat S_{0i}-S_{0i}|^2\mid T_i=1]^{1/2}
 \nonumber &\leq
   \E[|\widehat S_{2i}(\widehat S_{1i}-S_{1i})|^2
                                      \mid T_i=1]^{1/2}
 +
   \E[|S_{1i}(\widehat S_{2i}-S_{2i})|^2
                                      \mid T_i=1]^{1/2}
 \nonumber\\
 &\leq\frac{2c_1}{c_2}
     \sum_{j=1}^2
       \E[|\widehat S_{ji}-S_{ji}|^2\mid T_i=1]^{1/2}
 \nonumber\\
 &\leq c n^{-1/4}\delta_n.
 \label{eq:product-surrogate-ms-bound}
\end{align}
Here, the second inequality uses the uniform bounds on the multiplying
factors, and the third uses Equation~\eqref{eq:surrogate-ms-bound}.

\paragraph{STEP2-2: Bounding correction term $C_{i}$} For the correction term, write
\begin{align*}
 A_i&:=S_{1i}+S_{2i}-2S_{0i},
 &\widehat A_i&:=\widehat S_{1i}+\widehat S_{2i}-2\widehat S_{0i},\\
 B_i&:=\frac{\widetilde W_{3i}-m_1(1,\bX_i)}
                  {m_1(1,\bX_i)-m_1(0,\bX_i)},
 &\widehat B_i&:=\frac{\widetilde W_{3i}-\widehat m_1(1,\bX_i)}
                  {\widehat m_1(1,\bX_i)-\widehat m_1(0,\bX_i)}.
\end{align*}
Then $C_i=A_iB_i$ and $\widehat C_i=\widehat A_i\widehat B_i$.
By the triangle inequality, the bounds already established imply
\begin{align*}
 |A_i|
 &\leq |S_{1i}|+|S_{2i}|+2|S_{1i}S_{2i}| \leq\frac{4c_1}{c_2}+\frac{8c_1^2}{c_2^2},\\
 |B_i|
 &\leq\frac{|\widetilde W_{3i}|+|m_1(1,\bX_i)|}
                 {|m_1(1,\bX_i)-m_1(0,\bX_i)|}\leq\frac{2c_1}{c_2}.
\end{align*}
The same bounds hold for $\widehat A_i$ and $\widehat B_i$.
Also,
\begin{align*}
 \widehat A_i-A_i
 &=(\widehat S_{1i}-S_{1i})+(\widehat S_{2i}-S_{2i})
                       -2(\widehat S_{0i}-S_{0i}).
\end{align*}
Applying the triangle inequality gives
\begin{align}
\E[|\widehat A_i-A_i|^2\mid T_i=1]^{1/2}
&\leq\sum_{j=1}^2
      \E[|\widehat S_{ji}-S_{ji}|^2\mid T_i=1]^{1/2}
      +2\E[|\widehat S_{0i}-S_{0i}|^2\mid T_i=1]^{1/2}
 \nonumber\\
 &\leq2\sqrt{\frac{1+c_1^2}{c_2}}\,n^{-1/4}\delta_n
      +2\E[|\widehat S_{0i}-S_{0i}|^2\mid T_i=1]^{1/2}
 \nonumber\\
 &\leq c n^{-1/4}\delta_n.
 \label{eq:proxy-factor-ms-bound}
\end{align}
The last inequality follows by substituting the preceding surrogate
bounds. To compare the third-check factors, first add and subtract
the fraction with numerator $\widetilde W_{3i}-m_1(1,\bX_i)$
and the fitted denominator:
\begin{align*}
 \widehat B_i-B_i
 &=\frac{\widetilde W_{3i}-\widehat m_1(1,\bX_i)}
          {\widehat m_1(1,\bX_i)-\widehat m_1(0,\bX_i)}
   -\frac{\widetilde W_{3i}-m_1(1,\bX_i)}
          {m_1(1,\bX_i)-m_1(0,\bX_i)}\\
 &=-\frac{\widehat m_1(1,\bX_i)-m_1(1,\bX_i)}
          {\widehat m_1(1,\bX_i)-\widehat m_1(0,\bX_i)}\\
 &\quad+\{\widetilde W_{3i}-m_1(1,\bX_i)\}
   \left\{\frac1{\widehat m_1(1,\bX_i)-\widehat m_1(0,\bX_i)}
             -\frac1{m_1(1,\bX_i)-m_1(0,\bX_i)}\right\}\\
 &=-\frac{\widehat m_1(1,\bX_i)-m_1(1,\bX_i)}
          {\widehat m_1(1,\bX_i)-\widehat m_1(0,\bX_i)}\\
 &\quad-
   \frac{\widetilde W_{3i}-m_1(1,\bX_i)}
     {\{\widehat m_1(1,\bX_i)-\widehat m_1(0,\bX_i)\}
      \{m_1(1,\bX_i)-m_1(0,\bX_i)\}}\\
 &\qquad\times
   [\{\widehat m_1(1,\bX_i)-m_1(1,\bX_i)\}
        -\{\widehat m_1(0,\bX_i)-m_1(0,\bX_i)\}].
\end{align*}
Therefore, using the boundedness in Assumption~\ref{cond:inference}(a)--(b), we have
\begin{align*}
 |\widehat B_i-B_i|
 &\leq\frac1{c_2}|\widehat m_1(1,\bX_i)-m_1(1,\bX_i)|+\frac{2c_1}{c_2^2}
      |\{\widehat m_1(1,\bX_i)-m_1(1,\bX_i)\}
             -\{\widehat m_1(0,\bX_i)-m_1(0,\bX_i)\}|\\
 &\leq\left(\frac1{c_2}+\frac{2c_1}{c_2^2}\right)
        |\widehat m_1(1,\bX_i)-m_1(1,\bX_i)|+\frac{2c_1}{c_2^2}
        |\widehat m_1(0,\bX_i)-m_1(0,\bX_i)|.
\end{align*}
Taking second moments and applying the triangle inequality gives
\begin{align}
 &\E[|\widehat B_i-B_i|^2\mid T_i=1]^{1/2} \nonumber\\
 &\leq\left(\frac1{c_2}+\frac{2c_1}{c_2^2}\right)
     \E[|\widehat m_1(1,\bX_i)-m_1(1,\bX_i)|^2\mid T_i=1]^{1/2}
 +\frac{2c_1}{c_2^2}
     \E[|\widehat m_1(0,\bX_i)-m_1(0,\bX_i)|^2\mid T_i=1]^{1/2}
 \nonumber\\
 &\leq\frac1{\sqrt{c_2}}\left(\frac1{c_2}+\frac{2c_1}{c_2^2}\right)
     \E[|\widehat m_1(1,\bX_i)-m_1(1,\bX_i)|^2]^{1/2}
 +\frac{2c_1}{c_2^{5/2}}
     \E[|\widehat m_1(0,\bX_i)-m_1(0,\bX_i)|^2]^{1/2}
 \nonumber\\
 &\quad\leq c n^{-1/4}\delta_n,
 \label{eq:outcome-factor-ms-bound}
\end{align}
where the second inequality uses the covariate-moment bound at the beginning
of the proof, and the last inequality uses the rates for the two third-check means.

Adding and subtracting $A_i\widehat B_i$ gives
\begin{align*}
 \widehat C_i-C_i =\widehat A_i\widehat B_i-A_iB_i=\widehat A_i\widehat B_i-A_i\widehat B_i
                         +A_i\widehat B_i-A_iB_i=\widehat B_i(\widehat A_i-A_i)+A_i(\widehat B_i-B_i).
\end{align*}
We therefore obtain
\begin{align}
 &\E[|\widehat C_i-C_i|^2\mid T_i=1]^{1/2}
 \nonumber\\
 &\quad\leq
      \E[|\widehat B_i(\widehat A_i-A_i)|^2\mid T_i=1]^{1/2}
      +\E[|A_i(\widehat B_i-B_i)|^2\mid T_i=1]^{1/2}
 \nonumber\\
 &\quad\leq\frac{2c_1}{c_2}
       \E[|\widehat A_i-A_i|^2\mid T_i=1]^{1/2}
 +\left(\frac{4c_1}{c_2}+\frac{8c_1^2}{c_2^2}\right)
       \E[|\widehat B_i-B_i|^2\mid T_i=1]^{1/2}
 \nonumber\\
 &\quad\leq c n^{-1/4}\delta_n
 +\left(\frac{4c_1}{c_2}+\frac{8c_1^2}{c_2^2}\right)
       \E[|\widehat B_i-B_i|^2\mid T_i=1]^{1/2}
 \nonumber\\
 &\quad\leq c n^{-1/4}\delta_n.
 \label{eq:correction-ms-bound}
\end{align}
Here, the first inequality is the triangle inequality, and the second uses
boundedness of the multiplying factors. The third substitutes the bound
for $\widehat A_i-A_i$, and the fourth substitutes the bound for
$\widehat B_i-B_i$. Another application of the
triangle inequality yields
\begin{align}
 \E[|-\widehat S_{0i}+S_{0i}+\widehat C_i-C_i|^2
                                           \mid T_i=1]^{1/2}
 &\leq\E[|\widehat S_{0i}-S_{0i}|^2\mid T_i=1]^{1/2}
          +\E[|\widehat C_i-C_i|^2\mid T_i=1]^{1/2}
 \nonumber\\
 &\leq c n^{-1/4}\delta_n
          +\E[|\widehat C_i-C_i|^2\mid T_i=1]^{1/2}
 \nonumber\\
 &\leq c n^{-1/4}\delta_n.
 \label{eq:corrected-score-ms-bound}
\end{align}

\paragraph{STEP2-3: Bounding centered average of $H_i^Q$}

We can now bound the centered average of $H_i^Q$ in $L_2$ norm.
Applying the triangle inequality to its two terms gives
\begin{align}
 \E[(H_i^Q)^2]^{1/2}
 &\leq
   \E\!\left[\left(1-\frac{T_i}{\pi_1(\bX_i)}\right)^2
          |\widehat q(\bX_i)-q(\bX_i)|^2\right]^{1/2}
+
   \E\!\left[\frac{T_i^2}{\pi_1(\bX_i)^2}
       |-\widehat S_{0i}+S_{0i}+\widehat C_i-C_i|^2\right]^{1/2}
 \nonumber\\
 &\leq\frac1{\sqrt{c_2}}
       \E[|\widehat q(\bX_i)-q(\bX_i)|^2]^{1/2}
 +\frac1{c_2}
       \E[|-\widehat S_{0i}+S_{0i}+\widehat C_i-C_i|^2
                                         \mid T_i=1]^{1/2}
 \nonumber\\
 &\leq\frac1{\sqrt{c_2}}\delta_n
 +\frac1{c_2}
       \E[|-\widehat S_{0i}+S_{0i}+\widehat C_i-C_i|^2
                                         \mid T_i=1]^{1/2}
 \nonumber\\
 &\leq\frac1{\sqrt{c_2}}\delta_n
                    +c n^{-1/4}\delta_n
 \nonumber\\
 &\leq c\delta_n.
 \label{eq:probability-score-ms-bound}
\end{align}
For the second inequality, the first coefficient follows by conditioning
on $\bX_i$ and using the same Bernoulli second-moment calculation as
for $H_{ti}$, together with $\pi_1(\bX_i)\geq c_2$.
For the second coefficient, we use $T_i^2=T_i$ and
\begin{align*}
 \E\!\left[\frac{T_i}{\pi_1(\bX_i)^2}
       |-\widehat S_{0i}+S_{0i}+\widehat C_i-C_i|^2\right]
 &\leq\frac{\Pr(T_i=1)}{c_2^2}
       \E[|-\widehat S_{0i}+S_{0i}+\widehat C_i-C_i|^2
                                         \mid T_i=1]\\
 &\leq\frac1{c_2^2}
       \E[|-\widehat S_{0i}+S_{0i}+\widehat C_i-C_i|^2
                                         \mid T_i=1].
\end{align*}
Within an evaluation fold, the centered variables
$H_i^Q-\E[H_i^Q]$ are conditionally independent. Hence,
\begin{align*}
 \E\!\left[
    \left\{\frac{\sqrt n}{n_k}\sum_{i\in\mathcal I_k}
       (H_i^Q-\E[H_i^Q])\right\}^2\right]&=\frac{n}{n_k}\E[(H_i^Q-\E[H_i^Q])^2] \leq\frac{n}{n_k}\E[(H_i^Q)^2]\leq c\frac{n}{n_k}\delta_n^2.
\end{align*}
Conditional Chebyshev's inequality therefore bounds the probability
that the absolute value of this scaled average exceeds
$\varepsilon$ by $c(n/n_k)\delta_n^2/\varepsilon^2$ on the rate event.
This bound tends to zero, and the complement of the event has
vanishing probability. Averaging over the fixed number of folds gives
\begin{align}
 \sum_{k=1}^K\frac{n_k}{n}
   \left\{\frac1{n_k}\sum_{i\in\mathcal I_k}H_i^{Q,(-k)}
                      -\E[H_i^{Q,(-k)}]\right\}
 &=o_p(n^{-1/2}).
 \label{eq:centered-probability-bound}
\end{align}

\paragraph{STEP2-4: Bounding the conditional population remainder
$\E[H_i^Q]$}
It remains to bound the conditional population remainder
$\E[H_i^Q]$. The contribution from the fitted engagement
probability has expectation
\begin{align*}
 \E\!\left[
   \left(1-\frac{T_i}{\pi_1(\bX_i)}\right)
   \{\widehat q(\bX_i)-q(\bX_i)\}
 \right]
 &=
 \E\!\left[
   \{\widehat q(\bX_i)-q(\bX_i)\}
   \E\!\left[1-\frac{T_i}{\pi_1(\bX_i)}\,\middle|\,\bX_i\right]
 \right]\\
 &=
 \E\!\left[
   \{\widehat q(\bX_i)-q(\bX_i)\}
   \left\{1-\frac{\Pr(T_i=1\mid \bX_i)}{\pi_1(\bX_i)}\right\}
 \right]\\
 &=
 \E\!\left[
   \{\widehat q(\bX_i)-q(\bX_i)\}
   \left(1-\frac{\pi_1(\bX_i)}{\pi_1(\bX_i)}\right)
 \right]\\
 &\quad=0.
\end{align*}
where the third equality uses randomization. Consequently,
\begin{align}
 \E[H_i^Q]
    &= \E\biggl[\left(1-\frac{T_i}{\pi_1(\bX_i)}\right)
       \{\widehat q(\bX_i)-q(\bX_i)\}
     +\frac{T_i}{\pi_1(\bX_i)}
       \{-\widehat S_{0i}+S_{0i}+\widehat C_i-C_i\}\biggr] \nonumber \\
 &=\E\!\left[
   \frac{T_i}{\pi_1(\bX_i)}
   \{-\widehat S_{0i}+S_{0i}+\widehat C_i-C_i\}
 \right] \nonumber\\
 &=\Pr(T_i=1)\E\!\left[
   \frac{-\widehat S_{0i}+S_{0i}+\widehat C_i-C_i}{\pi_1(\bX_i)}
   \,\middle|\,T_i=1\right].
 \label{eq:probability-bias}
\end{align}

We calculate this expectation by first conditioning on
$W_i=w$, $T_i=1$, and $\bX_i=\bx$. Calibration gives
\begin{align*}
 \E[S_{ji}\mid W_i=w,T_i=1,\bX_i=\bx]=1-w,
 \qquad j=1,2.
\end{align*}
Since $S_{ji}$ is a function of the $j$th check and $\bX_i$,
conditional independence of the first two checks implies
\begin{align*}
 \E[S_{0i}\mid W_i=w,T_i=1,\bX_i=\bx]
 &=\E[S_{1i}S_{2i}\mid W_i=w,T_i=1,\bX_i=\bx]\\
 &=\E[S_{1i}\mid W_i=w,T_i=1,\bX_i=\bx]
   \E[S_{2i}\mid W_i=w,T_i=1,\bX_i=\bx]\\
 &=(1-w)^2\\
 &=1-w.
\end{align*}
The third equality substitutes the calibration identities,
and the last equality uses $w\in\{0,1\}$. 

It thus follows that
\begin{align}
 &\E[A_i\mid W_i=w,T_i=1,\bX_i=\bx]
 \nonumber\\
 &\quad=
 \E[S_{1i}\mid W_i=w,T_i=1,\bX_i=\bx]
 +\E[S_{2i}\mid W_i=w,T_i=1,\bX_i=\bx]
 -2\E[S_{0i}\mid W_i=w,T_i=1,\bX_i=\bx]
 \nonumber\\
 &\quad=(1-w)+(1-w)-2(1-w)
 \nonumber\\
 &\quad=0.
 \label{eq:proxy-factor-calibration}
\end{align}

The difference between the fitted and true product surrogates
can be expanded as
\begin{align*}
 \widehat S_{0i}-S_{0i}
 &=\widehat S_{1i}\widehat S_{2i}-S_{1i}S_{2i}\\
 &=\{S_{1i}+(\widehat S_{1i}-S_{1i})\}
   \{S_{2i}+(\widehat S_{2i}-S_{2i})\}-S_{1i}S_{2i}\\
 &=(\widehat S_{1i}-S_{1i})S_{2i}
   +S_{1i}(\widehat S_{2i}-S_{2i})
   +(\widehat S_{1i}-S_{1i})(\widehat S_{2i}-S_{2i}).
\end{align*}
Taking the conditional expectation first gives
\begin{align*}
 \E[\widehat S_{0i}-S_{0i}\mid W_i=w,T_i=1,\bX_i=\bx]
 &\quad=
 \E[(\widehat S_{1i}-S_{1i})S_{2i}
       \mid W_i=w,T_i=1,\bX_i=\bx]\\
 &\qquad+
 \E[S_{1i}(\widehat S_{2i}-S_{2i})
       \mid W_i=w,T_i=1,\bX_i=\bx]\\
 &\qquad+
 \E[(\widehat S_{1i}-S_{1i})(\widehat S_{2i}-S_{2i})
       \mid W_i=w,T_i=1,\bX_i=\bx].
\end{align*}
Each fitted surrogate is evaluated with its training sample
held fixed, so it remains a function of its own check and
$\bX_i$. Conditional independence therefore gives
\begin{align*}
 &\E[\widehat S_{0i}-S_{0i}\mid W_i=w,T_i=1,\bX_i=\bx]\\
 &\quad=
 \E[\widehat S_{1i}-S_{1i}\mid W_i=w,T_i=1,\bX_i=\bx]
 \E[S_{2i}\mid W_i=w,T_i=1,\bX_i=\bx]\\
 &\qquad+
 \E[S_{1i}\mid W_i=w,T_i=1,\bX_i=\bx]
 \E[\widehat S_{2i}-S_{2i}\mid W_i=w,T_i=1,\bX_i=\bx]\\
 &\qquad+
 \E[(\widehat S_{1i}-S_{1i})(\widehat S_{2i}-S_{2i})
       \mid W_i=w,T_i=1,\bX_i=\bx].
\end{align*}
Substituting the calibration identities now yields
\begin{align}
 &\E[\widehat S_{0i}-S_{0i}\mid W_i=w,T_i=1,\bX_i=\bx]
 \nonumber\\
 &\quad=(1-w)
 \E[\widehat S_{1i}-S_{1i}\mid W_i=w,T_i=1,\bX_i=\bx]
 \nonumber\\
 &\qquad+(1-w)
 \E[\widehat S_{2i}-S_{2i}\mid W_i=w,T_i=1,\bX_i=\bx]+
 \E[(\widehat S_{1i}-S_{1i})(\widehat S_{2i}-S_{2i})
       \mid W_i=w,T_i=1,\bX_i=\bx]
 \nonumber\\
 &\quad=(1-w)\sum_{j=1}^2
 \E[\widehat S_{ji}-S_{ji}\mid W_i=w,T_i=1,\bX_i=\bx]
 +
 \E[(\widehat S_{1i}-S_{1i})(\widehat S_{2i}-S_{2i})
       \mid W_i=w,T_i=1,\bX_i=\bx].
 \label{eq:product-bias-identity}
\end{align}

For the third-check factor, adding and subtracting
$\widehat m_1(w,\bx)$ in the numerator gives
\begin{align*}
 \frac{m_1(w,\bx)-\widehat m_1(1,\bx)}
       {\widehat m_1(1,\bx)-\widehat m_1(0,\bx)}&=
 \frac{\widehat m_1(w,\bx)-\widehat m_1(1,\bx)
       -\{\widehat m_1(w,\bx)-m_1(w,\bx)\}}
      {\widehat m_1(1,\bx)-\widehat m_1(0,\bx)}\\
 &=
 \frac{\widehat m_1(w,\bx)-\widehat m_1(1,\bx)}
      {\widehat m_1(1,\bx)-\widehat m_1(0,\bx)}
 -\frac{\widehat m_1(w,\bx)-m_1(w,\bx)}
       {\widehat m_1(1,\bx)-\widehat m_1(0,\bx)}.
\end{align*}
Because $w$ is binary,
\begin{align*}
 \widehat m_1(w,\bx)-\widehat m_1(1,\bx)
 &=(1-w)\widehat m_1(0,\bx)
   +w\widehat m_1(1,\bx)-\widehat m_1(1,\bx)\\
 &=(1-w)\widehat m_1(0,\bx)
   +(w-1)\widehat m_1(1,\bx)\\
 &=(w-1)\{\widehat m_1(1,\bx)-\widehat m_1(0,\bx)\}.
\end{align*}
Substitution into the preceding fraction then gives
\begin{align*}
 \frac{m_1(w,\bx)-\widehat m_1(1,\bx)}
       {\widehat m_1(1,\bx)-\widehat m_1(0,\bx)} &=
 \frac{(w-1)\{\widehat m_1(1,\bx)-\widehat m_1(0,\bx)\}}
      {\widehat m_1(1,\bx)-\widehat m_1(0,\bx)}
 -\frac{\widehat m_1(w,\bx)-m_1(w,\bx)}
       {\widehat m_1(1,\bx)-\widehat m_1(0,\bx)}\\
 &=(w-1)
 -\frac{\widehat m_1(w,\bx)-m_1(w,\bx)}
       {\widehat m_1(1,\bx)-\widehat m_1(0,\bx)}.
\end{align*}
The same binary identity applied to the true means gives
\begin{align*}
 \frac{m_1(w,\bx)-m_1(1,\bx)}{m_1(1,\bx)-m_1(0,\bx)}
 &=\frac{(w-1)\{m_1(1,\bx)-m_1(0,\bx)\}}
        {m_1(1,\bx)-m_1(0,\bx)} =w-1.
\end{align*}

We next calculate the mean of the correction difference.
First, its product representation implies
\begin{align*}
 &\E[\widehat C_i-C_i\mid W_i=w,T_i=1,\bX_i=\bx]\\
 &\quad=
 \E[\widehat B_i\widehat A_i\mid W_i=w,T_i=1,\bX_i=\bx]
 -\E[B_iA_i\mid W_i=w,T_i=1,\bX_i=\bx].
\end{align*}
Conditional independence of the third check from the first
two checks gives
\begin{align*}
 &\E[\widehat C_i-C_i\mid W_i=w,T_i=1,\bX_i=\bx]\\
 &\quad=
 \E[\widehat B_i\mid W_i=w,T_i=1,\bX_i=\bx]
 \E[\widehat A_i\mid W_i=w,T_i=1,\bX_i=\bx]\\
 &\qquad-
 \E[B_i\mid W_i=w,T_i=1,\bX_i=\bx]
 \E[A_i\mid W_i=w,T_i=1,\bX_i=\bx].
\end{align*}
Using the conditional mean of the third check yields
\begin{align*}
 &\E[\widehat C_i-C_i\mid W_i=w,T_i=1,\bX_i=\bx]\\
 &\quad=
 \frac{m_1(w,\bx)-\widehat m_1(1,\bx)}
      {\widehat m_1(1,\bx)-\widehat m_1(0,\bx)}
 \E[\widehat A_i\mid W_i=w,T_i=1,\bX_i=\bx]\\
 &\qquad-
 \frac{m_1(w,\bx)-m_1(1,\bx)}{m_1(1,\bx)-m_1(0,\bx)}
 \E[A_i\mid W_i=w,T_i=1,\bX_i=\bx]\\
 &\quad=
 \frac{m_1(w,\bx)-\widehat m_1(1,\bx)}
      {\widehat m_1(1,\bx)-\widehat m_1(0,\bx)}
 \E[\widehat A_i\mid W_i=w,T_i=1,\bX_i=\bx]\\
 &\qquad-(w-1)\E[A_i\mid W_i=w,T_i=1,\bX_i=\bx].
\end{align*}
The last equality uses the binary identity for the true
third-check means. Equation~\eqref{eq:proxy-factor-calibration}
then removes the last term:
\begin{align*}
 &\E[\widehat C_i-C_i\mid W_i=w,T_i=1,\bX_i=\bx]=
 \frac{m_1(w,\bx)-\widehat m_1(1,\bx)}
      {\widehat m_1(1,\bx)-\widehat m_1(0,\bx)}
 \E[\widehat A_i\mid W_i=w,T_i=1,\bX_i=\bx].
\end{align*}
Substituting the identity for the fitted fraction gives
\begin{align*}
 &\E[\widehat C_i-C_i\mid W_i=w,T_i=1,\bX_i=\bx]\\
 &\quad=
 \left\{(w-1)
 -\frac{\widehat m_1(w,\bx)-m_1(w,\bx)}
       {\widehat m_1(1,\bx)-\widehat m_1(0,\bx)}\right\}
 \E[\widehat A_i\mid W_i=w,T_i=1,\bX_i=\bx]\\
 &\quad=(w-1)\E[\widehat A_i\mid W_i=w,T_i=1,\bX_i=\bx] -
 \frac{\widehat m_1(w,\bx)-m_1(w,\bx)}
      {\widehat m_1(1,\bx)-\widehat m_1(0,\bx)}
 \E[\widehat A_i\mid W_i=w,T_i=1,\bX_i=\bx].
\end{align*}
Since the conditional mean of $A_i$ is zero, this is also
\begin{align}
 &\E[\widehat C_i-C_i\mid W_i=w,T_i=1,\bX_i=\bx]
 \nonumber\\
 &\quad=(w-1)
 \E[\widehat A_i-A_i\mid W_i=w,T_i=1,\bX_i=\bx]
 \nonumber\\
 &\qquad-
 \frac{\widehat m_1(w,\bx)-m_1(w,\bx)}
      {\widehat m_1(1,\bx)-\widehat m_1(0,\bx)}
 \E[\widehat A_i-A_i\mid W_i=w,T_i=1,\bX_i=\bx].
 \label{eq:correction-bias-identity}
\end{align}

We can now isolate the first-order terms in the remainder.
Substituting Equation~\eqref{eq:correction-bias-identity}
first gives
\begin{align*}
 &\E[-\widehat S_{0i}+S_{0i}+\widehat C_i-C_i
       \mid W_i=w,T_i=1,\bX_i=\bx]\\
 &\quad=-\E[\widehat S_{0i}-S_{0i}\mid W_i=w,T_i=1,\bX_i=\bx]\\
 &\qquad+(w-1)
 \E[\widehat A_i-A_i\mid W_i=w,T_i=1,\bX_i=\bx]\\
 &\qquad-
 \frac{\widehat m_1(w,\bx)-m_1(w,\bx)}
      {\widehat m_1(1,\bx)-\widehat m_1(0,\bx)}
 \E[\widehat A_i-A_i\mid W_i=w,T_i=1,\bX_i=\bx].
\end{align*}
The definition of $A_i$ implies
\begin{align*}
 \widehat A_i-A_i
 &=(\widehat S_{1i}-S_{1i})+(\widehat S_{2i}-S_{2i})
   -2(\widehat S_{0i}-S_{0i}).
\end{align*}
Substituting this expression into the term multiplied by
$w-1$ yields
\begin{align*}
 &\E[-\widehat S_{0i}+S_{0i}+\widehat C_i-C_i
       \mid W_i=w,T_i=1,\bX_i=\bx]\\
 &\quad=-\E[\widehat S_{0i}-S_{0i}\mid W_i=w,T_i=1,\bX_i=\bx]\\
 &\qquad+(w-1)\sum_{j=1}^2
 \E[\widehat S_{ji}-S_{ji}\mid W_i=w,T_i=1,\bX_i=\bx]\\
 &\qquad-2(w-1)
 \E[\widehat S_{0i}-S_{0i}\mid W_i=w,T_i=1,\bX_i=\bx]\\
 &\qquad-
 \frac{\widehat m_1(w,\bx)-m_1(w,\bx)}
      {\widehat m_1(1,\bx)-\widehat m_1(0,\bx)}
 \E[\widehat A_i-A_i\mid W_i=w,T_i=1,\bX_i=\bx]\\
 &\quad=(w-1)\sum_{j=1}^2
 \E[\widehat S_{ji}-S_{ji}\mid W_i=w,T_i=1,\bX_i=\bx]\\
 &\qquad+(1-2w)
 \E[\widehat S_{0i}-S_{0i}\mid W_i=w,T_i=1,\bX_i=\bx]\\
 &\qquad-
 \frac{\widehat m_1(w,\bx)-m_1(w,\bx)}
      {\widehat m_1(1,\bx)-\widehat m_1(0,\bx)}
 \E[\widehat A_i-A_i\mid W_i=w,T_i=1,\bX_i=\bx].
\end{align*}
Here, the last equality collects the two coefficients on
the product-surrogate error, using $-1-2(w-1)=1-2w$.
Next, Equation~\eqref{eq:product-bias-identity} gives
\begin{align*}
 &\E[-\widehat S_{0i}+S_{0i}+\widehat C_i-C_i
       \mid W_i=w,T_i=1,\bX_i=\bx]\\
 &\quad=(w-1)\sum_{j=1}^2
 \E[\widehat S_{ji}-S_{ji}\mid W_i=w,T_i=1,\bX_i=\bx]\\
 &\qquad+(1-2w)(1-w)\sum_{j=1}^2
 \E[\widehat S_{ji}-S_{ji}\mid W_i=w,T_i=1,\bX_i=\bx]\\
 &\qquad+(1-2w)
 \E[(\widehat S_{1i}-S_{1i})(\widehat S_{2i}-S_{2i})
       \mid W_i=w,T_i=1,\bX_i=\bx]\\
 &\qquad-
 \frac{\widehat m_1(w,\bx)-m_1(w,\bx)}
      {\widehat m_1(1,\bx)-\widehat m_1(0,\bx)}
 \E[\widehat A_i-A_i\mid W_i=w,T_i=1,\bX_i=\bx]\\
 &\quad=\{(w-1)+(1-2w)(1-w)\}\sum_{j=1}^2
 \E[\widehat S_{ji}-S_{ji}\mid W_i=w,T_i=1,\bX_i=\bx]\\
 &\qquad+(1-2w)
 \E[(\widehat S_{1i}-S_{1i})(\widehat S_{2i}-S_{2i})
       \mid W_i=w,T_i=1,\bX_i=\bx]\\
 &\qquad-
 \frac{\widehat m_1(w,\bx)-m_1(w,\bx)}
      {\widehat m_1(1,\bx)-\widehat m_1(0,\bx)}
 \E[\widehat A_i-A_i\mid W_i=w,T_i=1,\bX_i=\bx].
\end{align*}
The coefficient on the first-order surrogate errors is
\begin{align*}
 (w-1)+(1-2w)(1-w)
 &=w-1+1-3w+2w^2\\
 &=2w^2-2w\\
 &=-2w(1-w)\\
 &=0,
\end{align*}
where the last equality uses $w\in\{0,1\}$. Thus,
\begin{align*}
 &\E[-\widehat S_{0i}+S_{0i}+\widehat C_i-C_i
       \mid W_i=w,T_i=1,\bX_i=\bx]\\
 &\quad=(1-2w)
 \E[(\widehat S_{1i}-S_{1i})(\widehat S_{2i}-S_{2i})
       \mid W_i=w,T_i=1,\bX_i=\bx]\\
 &\qquad-
 \frac{\widehat m_1(w,\bx)-m_1(w,\bx)}
      {\widehat m_1(1,\bx)-\widehat m_1(0,\bx)}
 \E[\widehat A_i-A_i\mid W_i=w,T_i=1,\bX_i=\bx].
\end{align*}

Returning to Equation~\eqref{eq:probability-bias}, iterated
expectations give
\begin{align*}
 \E[H_i^Q]
 &=\Pr(T_i=1)\E\!\left[
   \frac1{\pi_1(\bX_i)}
   \E[-\widehat S_{0i}+S_{0i}+\widehat C_i-C_i
        \mid W_i,T_i=1,\bX_i]
   \,\middle|\,T_i=1
 \right]\\
 &=\Pr(T_i=1)\E\!\left[
   \frac{1-2W_i}{\pi_1(\bX_i)}
   \E[(\widehat S_{1i}-S_{1i})(\widehat S_{2i}-S_{2i})
        \mid W_i,T_i=1,\bX_i]
   \,\middle|\,T_i=1
 \right]\\
 &\quad-\Pr(T_i=1)\E\!\left[
   \frac{\widehat m_1(W_i,\bX_i)-m_1(W_i,\bX_i)}
        {\pi_1(\bX_i)\{\widehat m_1(1,\bX_i)-\widehat m_1(0,\bX_i)\}}
   \E[\widehat A_i-A_i\mid W_i,T_i=1,\bX_i]
   \,\middle|\,T_i=1
 \right].
\end{align*}
Both multiplying coefficients are functions of $(W_i,\bX_i)$.
Applying iterated expectations to the first term and then
to the second term gives
\begin{align}
 \E[H_i^Q]
 &=\Pr(T_i=1)\E\!\left[
   \frac{1-2W_i}{\pi_1(\bX_i)}(\widehat S_{1i}-S_{1i})(\widehat S_{2i}-S_{2i})
   \,\middle|\,T_i=1
 \right]
 \nonumber\\
 &\quad-\Pr(T_i=1)\E\!\left[
   \frac{\widehat m_1(W_i,\bX_i)-m_1(W_i,\bX_i)}
        {\pi_1(\bX_i)\{\widehat m_1(1,\bX_i)-\widehat m_1(0,\bX_i)\}}
   \E[\widehat A_i-A_i\mid W_i,T_i=1,\bX_i]
   \,\middle|\,T_i=1
 \right]
 \nonumber\\
 &=\Pr(T_i=1)\E\!\left[
   \frac{1-2W_i}{\pi_1(\bX_i)}(\widehat S_{1i}-S_{1i})(\widehat S_{2i}-S_{2i})
   \,\middle|\,T_i=1
 \right]
 \nonumber\\
 &\quad-\Pr(T_i=1)\E\!\left[
   \frac{\widehat m_1(W_i,\bX_i)-m_1(W_i,\bX_i)}
        {\pi_1(\bX_i)\{\widehat m_1(1,\bX_i)-\widehat m_1(0,\bX_i)\}}
   (\widehat A_i-A_i)
   \,\middle|\,T_i=1
 \right].
 \label{eq:exact-probability-remainder}
\end{align}

To bound the second product, we first control the error in
the third-check mean evaluated at the realized latent state.
Conditioning on $\bX_i$ within the treated arm gives
\begin{align*}
 &\E\!\left[
   |\widehat m_1(W_i,\bX_i)-m_1(W_i,\bX_i)|^2
   \,\middle|\,T_i=1
 \right]\\
 &\quad=\E\!\left[
   \E\!\left[
     |\widehat m_1(W_i,\bX_i)-m_1(W_i,\bX_i)|^2
     \,\middle|\,\bX_i,T_i=1
   \right]
   \,\middle|\,T_i=1
 \right]\\
 &\quad=\E\!\left[
   \{1-q(\bX_i)\}|\widehat m_1(0,\bX_i)-m_1(0,\bX_i)|^2
   +q(\bX_i)|\widehat m_1(1,\bX_i)-m_1(1,\bX_i)|^2
   \,\middle|\,T_i=1
 \right].
\end{align*}
Using the covariate-moment bound at the beginning of the proof, we obtain
\begin{align*}
 &\E\!\left[
   |\widehat m_1(W_i,\bX_i)-m_1(W_i,\bX_i)|^2
   \,\middle|\,T_i=1
 \right]\\
 &\quad\leq\frac1{c_2}\E\!\left[
   \{1-q(\bX_i)\}|\widehat m_1(0,\bX_i)-m_1(0,\bX_i)|^2
   +q(\bX_i)|\widehat m_1(1,\bX_i)-m_1(1,\bX_i)|^2
 \right]\\
 &\quad\leq\frac1{c_2}\E\!\left[
   |\widehat m_1(0,\bX_i)-m_1(0,\bX_i)|^2
   +|\widehat m_1(1,\bX_i)-m_1(1,\bX_i)|^2
 \right]\\
 &\quad=\frac1{c_2}\sum_{w=0}^1\E\!\left[
   |\widehat m_1(w,\bX_i)-m_1(w,\bX_i)|^2
 \right]\\
 &\quad\leq\frac2{c_2}n^{-1/2}\delta_n^2.
\end{align*}
The second inequality uses $0\leq q(\bX_i)\leq1$, and the
last inequality uses Assumption~\ref{cond:inference}(c).
Taking square roots yields
\begin{align*}
 &\E\!\left[
   |\widehat m_1(W_i,\bX_i)-m_1(W_i,\bX_i)|^2
   \,\middle|\,T_i=1
 \right]^{1/2}
 \leq\sqrt{\frac2{c_2}}\,n^{-1/4}\delta_n.
\end{align*}

Applying the triangle inequality to
Equation~\eqref{eq:exact-probability-remainder} gives
\begin{align*}
 |\E[H_i^Q]|
 &\leq
 \Pr(T_i=1)\left|\E\!\left[
   \frac{1-2W_i}{\pi_1(\bX_i)}(\widehat S_{1i}-S_{1i})(\widehat S_{2i}-S_{2i})
   \,\middle|\,T_i=1
 \right]\right|\\
 &\quad+
 \Pr(T_i=1)\left|\E\!\left[
   \frac{\widehat m_1(W_i,\bX_i)-m_1(W_i,\bX_i)}
        {\pi_1(\bX_i)\{\widehat m_1(1,\bX_i)-\widehat m_1(0,\bX_i)\}}
   (\widehat A_i-A_i)
   \,\middle|\,T_i=1
 \right]\right|\\
 &\leq\frac1{c_2}\E\!\left[
   |1-2W_i|\,|\widehat S_{1i}-S_{1i}|\,
   |\widehat S_{2i}-S_{2i}|
   \,\middle|\,T_i=1
 \right]\\
 &\quad+\frac1{c_2}\E\!\left[
   \frac{|\widehat m_1(W_i,\bX_i)-m_1(W_i,\bX_i)|}
        {|\widehat m_1(1,\bX_i)-\widehat m_1(0,\bX_i)|}
   |\widehat A_i-A_i|
   \,\middle|\,T_i=1
 \right].
\end{align*}
The second inequality uses $\Pr(T_i=1)\leq1$ and
$\pi_1(\bX_i)\geq c_2$.
Since $|1-2W_i|=1$, the first expectation becomes the
expectation of the product of the two surrogate errors:
\begin{align*}
 |\E[H_i^Q]|
 &\leq\frac1{c_2}\E\!\left[
   |\widehat S_{1i}-S_{1i}|\,|\widehat S_{2i}-S_{2i}|
   \,\middle|\,T_i=1
 \right]\\
 &\quad+\frac1{c_2}\E\!\left[
   \frac{|\widehat m_1(W_i,\bX_i)-m_1(W_i,\bX_i)|}
        {|\widehat m_1(1,\bX_i)-\widehat m_1(0,\bX_i)|}
   |\widehat A_i-A_i|
   \,\middle|\,T_i=1
 \right].
\end{align*}
The lower bound on the fitted third-check gap then gives
\begin{align*}
 |\E[H_i^Q]|
 &\leq\frac1{c_2}\E\!\left[
   |\widehat S_{1i}-S_{1i}|\,|\widehat S_{2i}-S_{2i}|
   \,\middle|\,T_i=1
 \right]\\
 &\quad+\frac1{c_2^2}\E\!\left[
   |\widehat m_1(W_i,\bX_i)-m_1(W_i,\bX_i)|\,
   |\widehat A_i-A_i|
   \,\middle|\,T_i=1
 \right].
\end{align*}
Applying the Cauchy--Schwarz inequality separately to the
two products yields
\begin{align*}
 |\E[H_i^Q]|
 &\leq
 \frac1{c_2}\E[|\widehat S_{1i}-S_{1i}|^2\mid T_i=1]^{1/2}
 \E[|\widehat S_{2i}-S_{2i}|^2\mid T_i=1]^{1/2}\\
 &\quad+\frac1{c_2^2}
 \E\!\left[
   |\widehat m_1(W_i,\bX_i)-m_1(W_i,\bX_i)|^2
   \,\middle|\,T_i=1
 \right]^{1/2}\\
 &\hspace{18mm}\times
 \E[|\widehat A_i-A_i|^2\mid T_i=1]^{1/2}.
\end{align*}
First inserting Equation~\eqref{eq:surrogate-ms-bound},
then the bound for the third-check mean, and finally
Equation~\eqref{eq:proxy-factor-ms-bound}, we obtain
\begin{align}
 |\E[H_i^Q]|
 &\leq
 \frac1{c_2}\left\{\sqrt{\frac{1+c_1^2}{c_2}}\,n^{-1/4}\delta_n\right\}^2
 +\frac1{c_2^2}
 \E\!\left[
   |\widehat m_1(W_i,\bX_i)-m_1(W_i,\bX_i)|^2
   \,\middle|\,T_i=1
 \right]^{1/2}
 \E[|\widehat A_i-A_i|^2\mid T_i=1]^{1/2}
 \nonumber\\
 &\leq
 \frac1{c_2}\left\{\sqrt{\frac{1+c_1^2}{c_2}}\,n^{-1/4}\delta_n\right\}^2
 +\frac{\sqrt{2/c_2}}{c_2^2}n^{-1/4}\delta_n
 \E[|\widehat A_i-A_i|^2\mid T_i=1]^{1/2}
 \nonumber\\
 &\leq
 \frac1{c_2}\left\{\sqrt{\frac{1+c_1^2}{c_2}}\,n^{-1/4}\delta_n\right\}^2
 +\frac{\sqrt{2/c_2}}{c_2^2}n^{-1/4}\delta_n
   (c n^{-1/4}\delta_n)
 \nonumber\\
 &=\left(\frac{1+c_1^2}{c_2^2}+\frac{\sqrt{2/c_2}\,c}{c_2^2}\right)
   n^{-1/2}\delta_n^2
 \nonumber\\
 &\leq c n^{-1/2}\delta_n^2
 \nonumber\\
 &=o(n^{-1/2}).
 \label{eq:probability-bias-bound}
\end{align}
The last equality uses $\delta_n\to0$. These bounds hold on
the training-sample event in
Assumption~\ref{cond:inference}(c). Since that event has
probability tending to one, each fold's conditional
population remainder is $o_p(n^{-1/2})$. With a fixed
number of folds,
\begin{align*}
 \sum_{k=1}^K\frac{n_k}{n}\E[H_i^{Q,(-k)}]
 =o_p(n^{-1/2}).
\end{align*}

\paragraph{STEP 3: Applying CLT} Substituting the centered empirical bound in
Equation~\eqref{eq:centered-probability-bound} into
Equation~\eqref{eq:probability-remainder-decomposition}
first gives
\begin{align*}
 \widehat Q-Q
 &=\frac1n\sum_{i=1}^n(\phi_i^Q-Q)
   +o_p(n^{-1/2})
   +\sum_{k=1}^K\frac{n_k}{n}\E[H_i^{Q,(-k)}].
\end{align*}
Substituting the population remainder bound then yields
\begin{align}
 \widehat Q-Q
 &=\frac1n\sum_{i=1}^n(\phi_i^Q-Q)
   +o_p(n^{-1/2})+o_p(n^{-1/2})
 =\frac1n\sum_{i=1}^n(\phi_i^Q-Q)+o_p(n^{-1/2}).
 \label{eq:probability-expansion}
\end{align}
Finally, we consider the ratio estimator. Proposition~\ref{prop:observed-if} proves that the uncentered influence function satisfies
\begin{align*}
 \E[\phi_{ti}-\Psi_t]&=0,\qquad t=0,1,\\
 \E[\phi_i^Q-Q]&=0.
\end{align*}
The outcomes and checks are bounded by
Assumption~\ref{cond:inference}(a). The denominators in the
calibrated surrogates and correction term are bounded away from zero by
Assumption~\ref{cond:inference}(b). Consequently, these surrogates
and the correction term are bounded. Since $\pi_t(\bX_i)\geq c_2$ for
$t=0,1$ by Assumption~\ref{cond:inference}(b), the component influence
functions have finite second moments. The central limit theorem therefore implies
\begin{align*}
 \frac1{\sqrt n}\sum_{i=1}^n(\phi_{ti}-\Psi_t)&=O_p(1),
 \qquad t=0,1,\\
 \frac1{\sqrt n}\sum_{i=1}^n(\phi_i^Q-Q)&=O_p(1).
\end{align*}
Equation~\eqref{eq:mean-expansion} then gives
$\widehat\Psi_t-\Psi_t=O_p(n^{-1/2})$ for $t=0,1$.
Equation~\eqref{eq:probability-expansion} gives
$\widehat Q-Q=O_p(n^{-1/2})$.

Assumption~\ref{cond:inference}(b) also implies
\begin{align*}
 Q&=\E[q(\bX_i)]\geq c_2.
\end{align*}
Thus, $\Pr(\widehat Q\geq Q/2)\to1$. On this event,
\begin{align*}
 \frac1{\widehat Q}-\frac1Q
 &=\frac{Q-\widehat Q}{Q\widehat Q}=-\frac{\widehat Q-Q}{Q\widehat Q}.
\end{align*}
The denominator is bounded away from zero on this event, so
\begin{align*}
 \frac1{\widehat Q}-\frac1Q=O_p(n^{-1/2}).
\end{align*}

Using the definition of $\widehat\Psi$, we obtain
\begin{align*}
 \widehat\Psi-\Psi
 &=\frac{\widehat\Psi_1-\widehat\Psi_0}{\widehat Q}-\Psi\\
 &=\frac{\widehat\Psi_1-\widehat\Psi_0-\Psi\widehat Q}
         {\widehat Q}\\
 &=\frac{(\widehat\Psi_1-\Psi_1)
          -(\widehat\Psi_0-\Psi_0)-\Psi(\widehat Q-Q)
          +(\Psi_1-\Psi_0-\Psi Q)}{\widehat Q}\\
 &=\frac{(\widehat\Psi_1-\Psi_1)
          -(\widehat\Psi_0-\Psi_0)-\Psi(\widehat Q-Q)}
         {\widehat Q}.
\end{align*}
The last equality uses $\Psi_1-\Psi_0=\Psi Q$.
Substituting Equation~\eqref{eq:mean-expansion} first gives
\begin{align*}
 \sqrt n(\widehat\Psi-\Psi)
 &=\frac1{\widehat Q}
   \left\{
    \frac1{\sqrt n}\sum_{i=1}^n
       \bigl[(\phi_{1i}-\Psi_1)-(\phi_{0i}-\Psi_0)\bigr]
       -\Psi\sqrt n(\widehat Q-Q)
   \right\}+o_p(1).
\end{align*}
Here, $1/\widehat Q=O_p(1)$ preserves the order of the remainder.
Next, substituting Equation~\eqref{eq:probability-expansion} yields
\begin{align*}
 \sqrt n(\widehat\Psi-\Psi)
 &=\frac1{\widehat Q\sqrt n}\sum_{i=1}^n
   \bigl[(\phi_{1i}-\Psi_1)-(\phi_{0i}-\Psi_0)
         -\Psi(\phi_i^Q-Q)\bigr]+o_p(1)\\
 &=\frac1{\widehat Q\sqrt n}\sum_{i=1}^n
   \bigl[\phi_{1i}-\phi_{0i}-\Psi\phi_i^Q
         -(\Psi_1-\Psi_0-\Psi Q)\bigr]+o_p(1)\\
 &=\frac1{\widehat Q\sqrt n}\sum_{i=1}^n
   (\phi_{1i}-\phi_{0i}-\Psi\phi_i^Q)+o_p(1).
\end{align*}
The last equality again uses $\Psi_1-\Psi_0=\Psi Q$.
Adding and subtracting the same sum divided by $Q$ gives
\begin{align*}
 \sqrt n(\widehat\Psi-\Psi)
 &=\frac1{Q\sqrt n}\sum_{i=1}^n
   (\phi_{1i}-\phi_{0i}-\Psi\phi_i^Q) +
   \left(\frac1{\widehat Q}-\frac1Q\right)
   \frac1{\sqrt n}\sum_{i=1}^n
   (\phi_{1i}-\phi_{0i}-\Psi\phi_i^Q)+o_p(1).
\end{align*}
The first factor in the product is $O_p(n^{-1/2})$.
For the second factor, the identity
\begin{align*}
 &\frac1{\sqrt n}\sum_{i=1}^n
   (\phi_{1i}-\phi_{0i}-\Psi\phi_i^Q)=
   \frac1{\sqrt n}\sum_{i=1}^n
   \bigl[(\phi_{1i}-\Psi_1)-(\phi_{0i}-\Psi_0)
         -\Psi(\phi_i^Q-Q)\bigr]
\end{align*}
shows that it is $O_p(1)$. Their product is therefore
$O_p(n^{-1/2})=o_p(1)$. It follows that
\begin{align*}
 \sqrt n(\widehat\Psi-\Psi)
 &=\frac1{Q\sqrt n}\sum_{i=1}^n
   (\phi_{1i}-\phi_{0i}-\Psi\phi_i^Q)+o_p(1)=\frac1{\sqrt n}\sum_{i=1}^n\phi_i+o_p(1),
\end{align*}
where the second equality uses Equation~\eqref{eq:ate-if}.

Finally,
\begin{align*}
 \E[\phi_i]
 &=\frac{\E[\phi_{1i}]-\E[\phi_{0i}]
              -\Psi\E[\phi_i^Q]}{Q}=\frac{\Psi_1-\Psi_0-\Psi Q}{Q} =0.
\end{align*}
The observations are independent and identically distributed, so the
$\phi_i$ are also independent and identically distributed.
Assumption~\ref{cond:inference}(d) gives
$0<\E[\phi_i^2]<\infty$. The central limit theorem therefore yields
\begin{align*}
 \sqrt n(\widehat\Psi-\Psi) = \frac1{\sqrt n}\sum_{i=1}^n\phi_i + o_p(1)
 \ \xrightarrow{d}\ \mathcal N\bigl(0,\E[\phi_i^2]\bigr).
\end{align*}
\end{proof}

\newpage
\subsection{Proof of Proposition~\ref{prop:covariance}}
\label{app:covariance}

\begin{proof}
Now, since $W_i$ is binary, we can write
\begin{align}
 \E[\widetilde W_{ji}\mid W_i,T_i=t,\bX_i=\bx]
 &=\E[\widetilde W_{ji}\mid W_i=0,T_i=t,\bX_i=\bx]
      +W_i d_t^j(\bx).
 \label{eq:affine-proxy}
\end{align}
By the conditional law of total covariance, we have
\begin{align*}
 &\Cov(\widetilde W_{ji},\widetilde W_{j^\prime,i}\mid T_i=t,\bX_i=\bx)\\
 &=\E[\Cov(\widetilde W_{ji}, \widetilde W_{j^\prime,i}\mid W_i,T_i=t,\bX_i=\bx)
                                      \mid T_i=t,\bX_i=\bx]\\
 &\quad+\Cov\!\left(
  \E[\widetilde W_{ji}\mid W_i,T_i=t,\bX_i=\bx],
  \E[\widetilde W_{j^\prime,i}\mid W_i,T_i=t,\bX_i=\bx]
  \,\middle|\,T_i=t,\bX_i=\bx\right).
\end{align*}
The first term is zero by conditional independence of the two
checks. We then apply Equation~\eqref{eq:affine-proxy} to each
check in the second term. Conditional on $(t,\bx)$, the intercepts
and the two separations are constants, leaving
\begin{align*}
\Cov(\widetilde W_{ji}, \widetilde W_{j^\prime,i}\mid T_i=t,\bX_i=\bx)
 &=\Cov\{W_id_t^j(\bx),W_i d_t^{j^\prime}(\bx)\mid T_i=t,\bX_i=\bx\}\\
 &=d_t^j(\bx)d_t^{j^\prime}(\bx)\V(W_i\mid T_i=t,\bX_i=\bx).
\end{align*}
This proves Equation~\eqref{eq:proxy-covariance} for each $(j,j^\prime)\in\{(1,2),(1,3),(2,3)\}$.
\end{proof}

\end{document}